\documentclass[preprint,pre,aps]{revtex4}
\usepackage[dvips]{graphicx}
\bibstyle{plain}
\begin{document}
\title{Mesoscopic theory for inhomogeneous mixtures}
\author{ A. Ciach}
\affiliation{Institute of Physical Chemistry,
 Polish Academy of Sciences, 01-224 Warszawa, Poland}
 \date{\today}

\begin{abstract}    
Mesoscopic density functional theory for inhomogeneous mixtures of sperical particles is developed in terms of 
mesoscopic volume fractions by a systematic coarse-graining procedure starting form microscopic theory. 
Approximate expressions for the correlation functions and for the grand potential are obtained for weak ordering on 
mesoscopic length scales. Stability analysis of the disordered phase
is performed in  mean-field approximation (MF) and beyond. MF
 shows existence of either a spinodal or a $\lambda$-surface on the volume-fractions - temperature phase diagram. 
Separation into homogeneous phases or formation of inhomogeneous distribution of particles 
 occurs on  the low-temperature side of the former or the latter surface respectively, depending on both 
the interaction potentials and the size ratios between  particles of different species.
Beyond MF the spinodal surface is shifted, and the instability at the  $\lambda$-surface is suppressed by fluctuations. We
interpret the  $\lambda$-surface as a borderline between homogeneous and inhomogeneous
 (containing clusters or other aggregates) structure of the disordered phase. For two-component systems 
explicit expressions for the MF spinodal and $\lambda$-surfaces are derived. Examples of 
interaction potentials of simple form are analyzed in some detail, in order to identify conditions leading to 
inhomogeneous structures. 

\end{abstract}

\maketitle
\section{Introduction}
One of the main problems that arise in theoretical description of complex fluids is
 the role of density fluctuations on the mesoscopic
length scale. Such fluctuations are not important in the case of
simple fluids, and for this reason simple liquids can be 
accurately described by the liquid theories \cite{hansen:76:0} which focus on the microscopic length scale, whereas
 the long-range fluctuations are treated via mean-field  (MF) approximation. 
An exception is the critical region where the long-range fluctuations  dominate. Universal
features of the critical phenomena are described by the phenomenological Landau-Ginzburg-Wilson (LGW) theory 
\cite{zinn-justin:89:0,amit:84:0}, 
because the field-theoretic methods allow for more accurate treatment of the dominant long-wavelength density 
fluctuations. In the LGW theory the microscopic structure is entirely 
neglected, however - the pair correlation function in the LGW theory decays monotonically.
 The two approaches (i) accurate description of the 
microscopic length scale and rough approximation for the long-wavelength fluctuations, and (ii)
 accurate description of the 
long-wavelength fluctuations with neglected microscopic structure are complementary. Both, used separately, 
give satisfactory description of simple fluids. Exact theories that are capable of description of 
nonuniversal features of phase transitions were also developed 
\cite{yukhnovskii:58:0,yukhnovskii:78:0,caillol:06:0,parola:85:0,parola:95:0,patsahan:99:0},
 but so far these theories were applied to homogeneous phases.

When there are competing tendencies in the interaction potentials, then self-assembly into different aggregates,
 living polymers, clusters,
 micelles or another objects ('supermolecules' having characteristic size) may occur. A notable example of such 
interactions is the effective
 short-range attraction long-range repulsion (SALR) potential
 \cite{pini:00:0,stradner:04:0,campbell:05:0,archer:07:0,archer:07:1,pini:06:0,imperio:04:0}. In
 addition to the liquid order on the microscopic length scale 
(described by the pair distribution function)
ordering on the mesoscopic length scale may be present in such systems. This additional ordering is associated with 
 packing of the 'supermolecules' in the lyotropic liquid crystalline phases. In the liquid theories, designed for 
description of the microscopic structure,
 the presence of such additional ordering is manifested by a lack of solutions of the associated equations
 \cite{pini:00:0,archer:07:0,archer:07:1,pini:06:0}.
 On the other hand, in the Landau theory  modified by Brazovskii the dominant
fluctuations of the order parameter (OP) are of finite wavelength~\cite{brazovskii:75:0}. 
The functional of the form postulated by Brazovskii was
used for a description of various amphiphilic systems 
\cite{leibler:80:0,teubner:87:0,fredrickson:87:0,gompper:94:0,podneks:96:0,ciach:01:2}, since the dominant
 finite-wavelength  fluctuations of the abstract order parameter (OP) can
 represent in particular the density fluctuations on the mesoscopic length scale. Indeed,  when the values
 of the phenomenological parameters in the functional are properly adjusted, the LB theory  predicts stability
 (or metastability) of lyotropic liquid 
crystalline phases observed in amphiphilic systems
\cite{leibler:80:0,fredrickson:87:0,gompper:94:0,podneks:96:0,ciach:01:2}. 
One should note that in the LB theory fluctuations of the OP lead to a change of the continuous transition between 
 disordered and lamellar phases
obtained in MF to weakly first order transition~\cite{brazovskii:75:0} which occurs at lower temperature. The
long-wavelength fluctuations in the LGW theory just modify the critical exponents, whereas the order of the transition 
remains the same.  The Landau-Brazovskii (LB) functional is quite general  (the OP can have different physical meaning), 
 and is expressed in terms of  phenomenological parameters (coupling constants)
whose precise relation to measurable quantities is not known - it should be derived from more fundamental microscopic 
theory. Because the LB theory correctly describes the qualitative properties of  systems self-assembling on the
 mesoscopic length scale, it is desirable to find a relation between 
the Brazovskii theory and the exact statistical mechanics. The main questions  are: (i) what 
 is the range of validity of the LGW and LB theories 
(ii) for what kind of interaction potentials the LB rather than the LGW theory is valid
(iii) how the phenomenological parameters are expressed in terms of thermodynamic parameters and
(effective) interaction potential. The approximate theory derived from the  statistical mechanics
 should allow for determination 
of phase diagrams and structure when the interaction potentials are known.

A natural theory to start with is the density functional theory (DFT) \cite{evans:79:0} which allows for description
 of inhomogeneous 
systems and in principle is exact.
 However, the exact form of the grand potential functional is not known. In the widely used versions of the DFT the 
contribution to the grand potential associated with interactions is of the
 MF type. This approximation works well for a description of the microscopic structure and away from the critical point. 
However,  more accurate approximation for the grand potential functional 
is desirable when the mesoscopic scale fluctuations dominate and may affect the order and location of the phase 
transition to a liquid crystalline phase. On the other hand, we have to make simplifying assumptions to make the
 theory tractable.

Since we need a description of the ordering on the mesoscopic
 length scale, we may introduce {\it mesoscopic density} that describes the distribution of particles less accurately 
than the microscopic density, but more accurately than the average density.  Such an approach was proposed in 
Ref.\cite{ciach:08:1} for a one component system of 
spherical particles.  The mesoscopic density is defined as the microscopic 
density averaged over regions larger than the  molecules and smaller than the characteristic length of ordering 
(for example, the size of the clusters). 
Precise definition is given for multicomponent systems in the next section. Probability of spontaneous appearance
 of particular mesoscopic density field was derived from the statistical 
mechanics by  integrating the probability distribution over all microscopic states under the constraint of fixed  
mesoscopic density field under consideration. This method is analogous to integrating the probability distribution over all microscopic
 states under the constraint of fixed average density in  macroscopic parts of the system. The only difference is that 
 the constraint imposed on the microscopic density 
has a form of the field which varies on the mesoscopic length scale. 
The grand potential functional of the mesoscopic density field  derived in Ref.\cite{ciach:08:1}
  consists of two terms.  The first term contains
 contributions from fluctuations on the microscopic length scale under the 
constraint of fixed 
mesoscopic density  $\bar\rho({\bf r})$. This term resembles standard DFT. The second contribution is associated 
with mesoscopic
 fluctuations $\phi({\bf r})$
that can occur in the system when the constraint $\bar\rho({\bf r})$ is removed.

In the MF approximation the contribution to the grand potential associated with mesoscopic-length scale fluctuations
 is neglected. In this version of MF the average density is approximated by the most probable mesoscopic
 density \cite{ciach:08:1}. However, in  parts of the 
phase diagram that are close to microphase separation the fluctuation contribution can be comparable to 
the first term, and the  average density can be significantly different from the most probable mesoscopic
 density. In such cases the fluctuations cannot be neglected. 

It is worthwhile to note that  the 
relation between the systems with and without the constraint on the mesoscopic density distribution resembles
 the relation between the canonical and the grand canonical ensembles. 
There is some loose analogy between the system in the presence of the mesoscopic constraint imposed on the microscopic
 density
 distribution, and a macroscopic system with fixed number of particles $N=N_0$, described by the canonical ensemble.
 When the constraint of compatibility
between the microscopic density distribution and the mesoscopic field $\bar\rho({\bf r})$ is removed 
and $\langle\phi({\bf r})\rangle=0$, then the system is analogous to the open system with fluctuating $N$ such
 that $ \langle N\rangle=N_0$ (grand canonical ensemble). 
The canonical and grand canonical ensembles with $N_0=\langle N\rangle$ are equivalent only
 far from phase transitions, when the fluctuations  are small, $\langle (N- N_0)^2\rangle \propto \chi_T N_0$. 
Close to phase transitions the compressibility $\chi_T$  is large (diverges at the transition), and fluctuations cannot be neglected.
At the phase coexistence the most probable density distribution in the open system
corresponds to either the gas $\rho_g$ or the liquid $\rho_l$ density in an absence of any 
constraints or external fields. However, in the grand canonical ensemble the ensemble average in 
an absence of any constraints or external fields yields a constant density
 $(\rho_g+\rho_l)/2$, although homogeneous microscopic states with such
density  occur with 
negligible probability. The difference between this case and microphase separation concerns the 
extent of the regions having different density (or composition) - mesoscopic rather than macroscopic parts of the
 system - and in turn the time scale associated with displacements of these regions, i.e. with the mesoscopic
  rather than macroscopic fluctuations. While it is justified to neglect macroscopic fluctuations in studies
 of  coexisting homogeneous phases, in the case of microphase separation the 
mesoscopic fluctuations influence the experimentally observed properties of the system. 

 The fluctuation contribution to the grand potential reduces to the form similar to the LB theory in the case of
 weak ordering on the length scale significantly larger than the molecular size \cite{ciach:08:1}. Such kind of
 ordering occurs in soft-matter systems, and the results of Ref. \cite{ciach:08:1} confirm validity 
of the LB theory for soft matter.
 The fluctuation contribution can be treated by field-theoretic methods, and in the theory developed in 
Ref.\cite{ciach:08:1} the DFT and field-theoretic
 methods are both used.

The theory developed in Ref.\cite{ciach:08:1} is restricted to a one-component system, whereas the soft-matter systems 
are usually multicomponent. The size of solvent molecules can be  several 
orders of magnitude smaller than the size of proteins,
 nanoparticles or colloids, and the solvent molecules can be taken into account only via 
solvent-mediated effective interactions between  solute particles \cite{dijkstra:99:0,dijkstra:00:0}. However, 
the effectively one-component system might lead to incorrect predictions when the 
size ratio is not very large, and  when mesoscopic fluctuations of the solvent are important.
 The purpose of this work is an extension of 
the mesoscopic DFT to the case of multicomponent systems of particles of arbitrary sizes. 

In sec.2  general framework of the theory for multicomponent
 systems is introduced by a systematic coarse-graining procedure.  Mesoscopic volume fractions are defined,
 and expressions for the grand potential and correlation
 functions are derived in the same section.
 In sec.3 approximate theory for weak ordering is developed, and the role of mesoscopic fluctuations for  
stability of the disordered phase is discussed. Two-component systems  of particles of different sizes  are 
studied in more detail in sec.4.  The MF theory is illustrated by three simple examples. 
Equations derived in sec.4 are of MF type, and provide information whether 
 inhomogeneities  on the mesoscopic length scale (clusters or soft crystals) 
are formed, or the system can phase separate into homogeneous phases  for given interaction potentials and size ratios. 
In future studies the theory can be applied to different inhomogeneous systems along the lines 
described in sec.3.
 
\section{Coarse graining}
\subsection{microscopic density and microscopic volume fraction}
We consider an $n$-component mixture of nearly spherical particles, with the components labeled by Greek letters. 
The diameter of the hard core of the particle of the specie $\alpha$ is denoted by $\sigma_{\alpha}$.
A microscopic state is defined by the positions of the centers of mass of $N_1, ..., N_n$ particles,
\begin{equation}
\label{micro}
{\cal M}= \{\{{\bf r}^{\alpha}_i\}_{i=1,...,N_{\alpha}},\alpha=1,...,n\}
\end{equation}
where ${\bf r}^{\alpha}_i$ denotes the position of the $i$-th particle of the $\alpha$-th specie, and $N_{\alpha}$ 
denotes the number of particles of the $\alpha$-th specie in the considered microstate. 
Microscopic density of the $\alpha$-th specie is given by the standard definition,
\begin{equation}
\label{micror}
 \hat\rho_{\alpha}({\bf r},{\cal M}):=\sum_{i=1}^{N_{\alpha}}\delta({\bf r}-{\bf r}_i^{\alpha}).
\end{equation}

For particles of different sizes, for example for a mixture of nanoparticles and small molecules, 
it is convenient to introduce microscopic density that takes into account distribution of matter inside the molecules, 
and instead of (\ref{micror}) we introduce
\begin{equation}
\label{microre}
\hat\zeta_{\alpha}({\bf r},{\cal M}):=\sum_{i=1}^{N_{\alpha}}f_{\alpha}(|{\bf r}-{\bf r}_i^{\alpha}|)
\end{equation}
where  spherically-symmetric structure of molecules is assumed,  $\int_{\bf r} f_{\alpha}(r)=v_{\alpha}$ with 
 $v_{\alpha}=\pi\sigma_{\alpha}^3/6 $ denoting the volume of the particle of the specie $\alpha$,  and $f_{\alpha}(r)$ 
describes distribution of matter inside such particle at the distance $r$ from its center.
 For brevity we shall use the notation $\int_{\bf r}\equiv \int d{\bf r}$, indicating the integration region $S$ 
by $\int_{{\bf r'}\in S}$ when necessary. For constant density inside the particle Eq.(\ref{microre})  reduces to 
the microscopic volume fraction defined by
\begin{equation}
\label{microe}
 \hat\zeta_{\alpha}({\bf r},{\cal M}):=
\sum_{i=1}^{N_{\alpha}}\theta\Big(\frac{\sigma_{\alpha}}{2}-|{\bf r}-{\bf r}_i^{\alpha}|\Big)
\end{equation}
where $\theta(r)$ is the Heaviside unit step function. Integration of $\hat\zeta_{\alpha}({\bf r},{\cal M})$ 
over the system volume gives the volume occupied by the particles.
The interaction energy for a pair of particles $i,j$ of the  $\alpha$, $\beta$ species
with the centers at ${\bf r}_i^{\alpha}$ and
 ${\bf r}_j^{\beta}$ respectively is
\begin{equation}
 U_{\alpha \beta}(|{\bf r}_i^{\alpha}-{\bf r}_j^{\beta}|)=\int_{\bf r}\int_{\bf r'}  
f_{\alpha}(|{\bf r}-{\bf r}_i^{\alpha}|) V_{\alpha \beta}({\bf r},{\bf r}') f_{\beta}(|{\bf r}'-{\bf r}_j^{\beta}|).
\end{equation} 
Summation convention for repeated Greek indexes is assumed above and in the whole article.
   $ V_{\alpha \beta}({\bf r},{\bf r}')d{\bf r}d{\bf r'}$ is the interaction potential between the infinitesimal volumes 
$d{\bf r}$ and $d{\bf r'}$ around the points ${\bf r}$ and ${\bf r'}$ inside the particles $\alpha$ and $\beta$. 
The energy of the system in the microstate defined by (\ref{microre}) or (\ref{micror})  can be written as
\begin{eqnarray}
 \label{emicroe}
E[{\cal M}]=\frac{1}{2}\int_{\bf r}\int_{\bf r'} \hat\zeta_{\alpha}({\bf r},{\cal M}) 
 V_{\alpha \beta}({\bf r},{\bf r}')\hat\zeta_{\beta}({\bf r}',{\cal M})\\\nonumber
=
\frac{1}{2}\sum_{\alpha=1}^n\sum_{\beta=1}^n\sum_{i=1}^{N_{\alpha}}\sum_{j=1}^{N_{\beta}} 
U_{\alpha \beta}(|{\bf r}_i^{\alpha}-{\bf r}_j^{\beta}|)\\
\nonumber
=\frac{1}{2}\int_{\bf r}\int_{\bf r'} \hat\rho_{\alpha}({\bf r},{\cal M}) 
 U_{\alpha \beta}({\bf r},{\bf r}')\hat\rho_{\beta}({\bf r}',{\cal M})
\end{eqnarray}

\subsection{Mesoscopic density and mesoscopic volume fraction}
Let us  choose the mesoscopic length scale $R$ and consider spheres $S_R({\bf r})$ of radius $R$ and centers 
at ${\bf r}$ that cover the whole volume $V$ of the system.
 In order to describe ordering on the length scale $\lambda$, we should choose $R< \lambda$. We define the 
{\it mesoscopic} density of the specie $\alpha$ by an extension of the  definition introduced for a one-component
 system in Ref.\cite{ciach:08:1} 
\begin{equation}
\label{mr}
 \rho_{\alpha}({\bf r}):=\frac{1}{V_S} \int_{{\bf r'}\in S_R({\bf r})}\hat\rho_{\alpha}({\bf r'},{\cal M}),
\end{equation}
where $V_S=4\pi R^3/3$ is the volume of the sphere $S_R({\bf r})$. Similarly, the {\it mesoscopic} volume fraction of the specie $\alpha$ at ${\bf r}$ is
 defined by
\begin{equation}
\label{me}
 \zeta_{\alpha}({\bf r}):=\frac{1}{V_S} \int_{{\bf r'}\in S_R({\bf r})}\hat\zeta_{\alpha}({\bf r'},{\cal M}).
\end{equation}
For an illustration, the mesoscopic density and the mesoscopic volume fraction are shown in Fig. 1
 for a one-component system, when a single particle is located at ${\bf r}={\bf 0}$, for three different mesoscopic 
length scales $R$. In this case the center of the particle is inside (outside) the sphere $S_R({\bf r})$ for
 $r<2R/\sigma_{\alpha}$ ($r>2R/\sigma_{\alpha}$), therefore for $r=2R/\sigma_{\alpha}$ the number density (\ref{mr})
 has a discontinuity (Fig.2).
For increasing length scale of coarse-graining, the difference between $\zeta$ and $\rho v$ decreases.

\begin{figure}
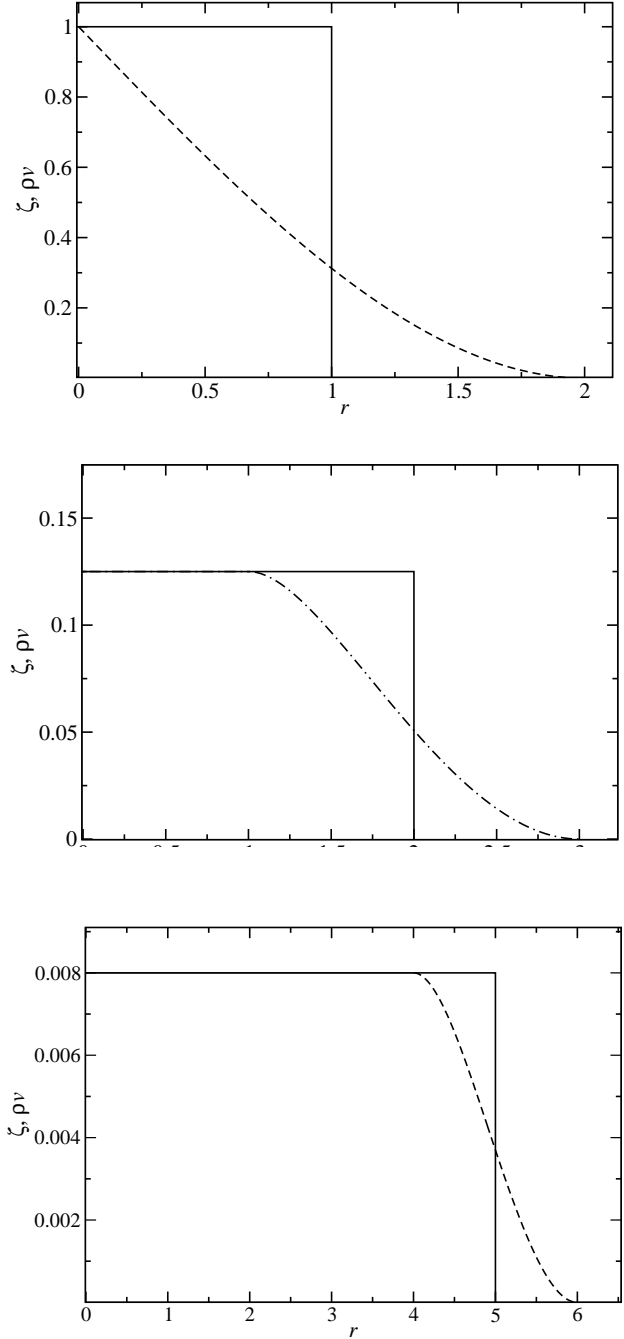

\includegraphics[scale=0.33]{fig1a.eps}\\
\vskip0.6cm
\includegraphics[scale=0.33]{fig1b.eps}\\
\vskip0.6cm
\includegraphics[scale=0.33]{fig1c.eps}

\caption{ One-component system  and the microstate in which a single hard sphere of a radius $\sigma/2$ is 
located at ${\bf r}={\bf 0}$ is considered for different length scales of coarse-graining. 
The mesoscopic volume-fraction $\zeta$ defined in Eq.(\ref{me}) is shown by the dashed lines, and $\rho v$
is shown by the solid lines. 
  $v=\pi\sigma^3/6$  and $\rho$ is defined in Eq.(\ref{mr}) (based on the standard definition of 
the microscopic density (\ref{micror})). Top panel: $2R/\sigma=1$. Central panel:
 $2R/\sigma=2$. Bottom panel: $2R/\sigma=5$. In this particularly simple case the fields (\ref{me})
 and (\ref{mr}) are functions of the distance $r$ from the center of the hard sphere. In each case
 $\int_{\bf r}\zeta({\bf r})=v$. The distance $r$ is in $\sigma/2$ units, $\zeta$ and $\rho v$ are dimensionless. 
}
\end{figure}
\begin{figure}
\includegraphics[scale=0.33]{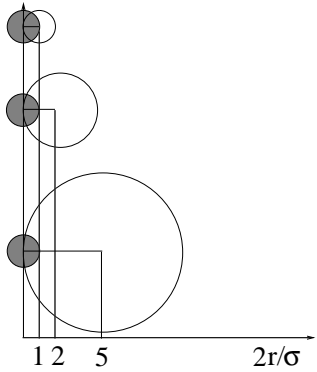}
\caption{Cartoon showing distances corresponding to the discontinuities of the mesoscopic number density for different scales 
of coarse graining. From top to bottom $2R/\sigma=1$, $2R/\sigma=2$ and $2R/\sigma=5$, as in Fig.1.
 Shaded circle represents the particle located at ${\bf r=0}$, open circles represent 
the spheres  $S_R({\bf r})$ with the centers at ${\bf r}$ over which the density or volume fraction is averaged. The mesoscopic density at the point ${\bf r}$ 
is the number of centers of particles inside the sphere $S_R({\bf r})$ divided by $4\pi R^3/3$.
 The mesoscopic volume fraction is the fraction  of the volume  $4\pi R^3/3$ that is occupied by the particles.}
\end{figure}
For a chosen length scale $R$ the mesostate can be defined by  $\{\zeta\}=\{\zeta_1({\bf r}), ... ,\zeta_n({\bf r})\}$
 or by $\{\rho\}=\{\rho_1({\bf r}), ... ,\rho_n({\bf r})\}$. 
Note that  Eq.(\ref{mr}) or  Eq.(\ref{me}) describes a constraint imposed on the microscopic states, and 
 $\{\zeta\}$ is equivalent to the constraint (\ref{me}) imposed on all the components.  
For a chosen length scale $R$ all microscopic states can be separated into disjoint subsets, such
 that the microstates belonging to a particular subset are compatible with the same constraint (Eq.(\ref{me})). 
Microstates belonging to different subsets are compatible with  different constraints, i.e. with a different form of 
 $\{\zeta\}$ or $\{\rho\}$.  

Probability  density of a spontaneous occurrence of the mesostate $\{\zeta\}$  is given by
\begin{equation}
\label{p2}
  p[\{\zeta\}]={\Xi}^{-1}e^{-\beta\Omega_{co}[\{\zeta\}]}
 \end{equation}
where 
\begin{equation}
\label{Xi2}
 \Xi=\int^{'} D \zeta_1 ... \int^{'} D \zeta_n e^{-\beta\Omega_{co}[\{\zeta\}]}.
\end{equation}
and
\begin{equation}
\label{Omcod}
 e^{-\beta\Omega_{co}[\{\zeta\}]}=\int_{{\cal M}\in\{\{\zeta\},R\}}
e^{-\beta(E[{\cal M}]-\int_{\bf r}\bar\mu_{\alpha}\zeta_{\alpha}({\bf r}))}.
\end{equation}
 $E[{\cal M}]$ is the microscopic Hamiltonian,   and  $ \int_{{\cal M}\in\{\{\zeta\},R\}}$ is a 
symbolic notation for the integration over all microstates compatible with $\{\zeta\}$ according to Eq.(\ref{me}). 
 $\bar\mu_{\alpha}=\mu_{\alpha}/v_{\alpha}$ and $T$ are the chemical potential of the specie $\alpha$ 
(in appropriate units) and temperature respectively, and $\beta=1/k_BT$, with $k_B$ denoting the Boltzmann constant.
 $\Omega_{co}[\{\zeta\}]$ 
is  the grand potential in the presence of the constraints  $\{\zeta\}$  (Eq.(\ref{me})) imposed on the system. 
The functional integral $ \int^{'} D\zeta_1 ... \int^{'} D \zeta_n$ in Eq.(\ref{Xi2}) is over all mesostates $\{\zeta\} $,
 which is indicated by the prime. In analogous way we can consider mesoscopic theory based on the mesoscopic density. 

We obtain a mesoscopic theory with the same structure as the standard statistical mechanics. 
The integration over all microstates is replaced in Eq.(\ref{Xi2}) by the integration over all mesostates. 
The Hamiltonian is replaced in Eq.(\ref{p2}) and (\ref{Xi2}) by the grand potential in the presence of the constraint of 
compatibility with the given mesostate that is imposed on the microstates. The above formulas are exact. 
So far we just rearranged the summation over microstates. The reason for doing so is the possibility of performing
 the summation over the mesostates and over the microstates compatible with a particular mesostate by different methods.

Grand potential in the presence of the mesoscopic constraint can be written in the form 
 \begin{equation}
\label{Omco}
\Omega_{co}=U-TS-\mu_{\alpha} N_{\alpha},
\end{equation}
 where $U,S$ and $ N_{\alpha}$ are the internal energy, entropy and the number of molecules of the specie $\alpha$ 
respectively in the system with the constraint  (\ref{me})  imposed on the microscopic densities. 
 $U$ is given by the  expression
\begin{equation}
\label{U}
 U[\{\zeta\}]=\frac{1}{2}\int_{\bf r_1}\int_{\bf r_2} V_{\alpha \beta}^{co}({\bf r}_1-{\bf r}_2)
\zeta_{\alpha}({\bf r}_1)\zeta_{\beta}({\bf r}_2)=
\frac{1}{2}\int_{\bf r_1}\int_{\bf r_2}U_{\alpha \beta}^{co}({\bf r}_1-{\bf r}_2)
\rho_{\alpha}({\bf r}_1)\rho_{\beta}({\bf r}_2),
\end{equation}
where 
\begin{eqnarray}
\label{Vco}
  V_{\alpha \beta}^{co}({\bf r}_1-{\bf r}_2)=   V_{\alpha \beta}(r_{12})
g_{\alpha \beta}^{\zeta co}({\bf r}_1-{\bf r}_2)
\end{eqnarray}
 $r_{12}=|{\bf r}_1-{\bf r}_2|$, and
\begin{eqnarray}
\label{gzco}
 g_{\alpha \beta}^{\zeta co}({\bf r}_1-{\bf r}_2)=\frac{\langle \hat \zeta_{\alpha}({\bf r}_1) 
\hat \zeta_{\beta}({\bf r}_2)\rangle}{\zeta_{\alpha}({\bf r}_1)\zeta_{\beta}({\bf r}_2)}
\end{eqnarray}
  is the microscopic pair correlation function for the volume fraction, in the presence of the constraint
 (\ref{me})  imposed on the microscopic states. $U_{\alpha \beta}^{co}({\bf r}_1-{\bf r}_2)$
is given by an expression analogous to Eq.(\ref{Vco}),  with $ g_{\alpha \beta}^{\zeta co}$ replaced by 
 $ g_{\alpha \beta}^{ co}$,
the standard pair correlation function in the presence of the constraint (\ref{mr}).
 Note that the above functions differ from each other. In particular, for  $ g_{\alpha \beta}^{\zeta co}$
 smooth increase from zero is expected for
 $r_{12}$ increasing from zero, whereas the correlation function for the microscopic density (Eq. (\ref{micror})) 
vanishes for 
$r_{12}<\sigma_{\alpha\beta}$. 

The advantage of $\zeta_{\alpha}({\bf r})$ is its continuity (see Fig.1). In addition,
 $\zeta_{\alpha}({\bf r})\le \zeta_{cp}$ for all ${\bf r}$, where $\zeta_{cp}$ is the close-packing 
volume fraction, and the gradient of $\zeta$ is small, $|\nabla\zeta|<1/R$. The disadvantage of 
 $\zeta_{\alpha}({\bf r})$ is the expression for the energy (\ref{U}) in terms of the pair correlation function
 for the volume fraction, Eq.(\ref{gzco}), which was not studied.
 The mesoscopic 
density (\ref{mr}) has discontinuities (see Fig.1), and for significantly different sizes of particles the 
number densities for different components may differ by several orders of magnitude.
 On the other hand, the expression for the energy  (\ref{U}) has a 
standard form in terms of the well known correlation function.
When the ordering occurs on the length scale significantly larger than the size of 
the particles,  we can make the approximation (see Fig.1)
\begin{eqnarray}
\label{zvr}
\zeta_{\alpha}({\bf r})\approx \rho_{\alpha}({\bf r}) v_{\alpha}.
\end{eqnarray}
Inserting the above expression for $ \rho_{\alpha}$ into Eq.(\ref{U}), yields 
\begin{eqnarray}
\label{VU}
  V^{co}_{\alpha \beta}({\bf r},{\bf r}')\approx\frac{U^{co}_{\alpha \beta}({\bf r},{\bf r}')}{v_{\alpha}v_{\beta}}.
\end{eqnarray}

It is important to remember that the approximation (\ref{VU}) is only valid when the ordering occurs on the length scale
significantly larger than the size of particles.
 
We further assume that  the entropy $S$ satisfies the relation $-TS=F_h$, where $F_h$ is the free-energy of 
the hard-sphere reference system with the constraint  (\ref{me}) imposed on the  microscopic volume fractions. 

\subsection{Grand-potential functional  and mesoscopic  correlation functions }
 Let us introduce  external fields $\{J\}=\{J_1({\bf r}),...,  J_n({\bf r})\} $ and the grand-thermodynamic
 potential functional 
\begin{equation}
\label{OmJ}
 \Omega[\{\beta J\}]:=-k_BT\log\Big[\int^{'}D\zeta_1 ...\int^{'}D\zeta_n   e^{-\beta[\Omega_{co}[\{\zeta\}]-
\int_{\bf r}J_{\alpha}({\bf r})\zeta_{\alpha}({\bf r})]}\Big].
\end{equation}
$-\beta\Omega[\beta J]$ is 
the generating functional for the (connected) correlation functions for the mesoscopic volume fractions,
\begin{equation}
\label{genfu}
 \langle\zeta_{\alpha_1}({\bf r}_1)...\zeta_{\alpha_n}({\bf r}_n)\rangle^{con}=
\frac{\delta^n(-\beta\Omega[\beta J])}{\delta(\beta J_{\alpha_1}({\bf r}_1))...\delta(\beta J_{\alpha_n}({\bf r}_n))}.
\end{equation}
 We introduce the notation
\begin{eqnarray}
\label{corfuecon}
G^{m}_{\alpha\beta}({\bf r}_1-{\bf r}_2)= \langle\zeta_{\alpha}({\bf r}_1)\zeta_{\beta}({\bf r}_2)\rangle^{con}=
\langle\zeta_{\alpha}({\bf r}_1)\zeta_{\beta}({\bf r}_2)\rangle-\langle\zeta_{\alpha}({\bf r}_1)
\rangle\langle\zeta_{\beta}({\bf r}_2)\rangle.
\end{eqnarray}
The relation between the mesoscopic and the microscopic correlation functions resulting from the definition
 of the mesoscopic volume fraction (\ref{me}) is given by
\begin{eqnarray}
\label{corfue}
 \langle\zeta_{\alpha_1}({\bf r}_1)\cdot ... \cdot \zeta_{\alpha_n}({\bf r}_n)\rangle\\
\nonumber
=
\frac{\int^{'}D\zeta_1 ... \int^{'}D\zeta_n  e^{-\beta[\Omega_{co}[\{\beta J\}]-
\int_{\bf r}J_{\alpha}({\bf r})\zeta_{\alpha}({\bf r})]}\zeta_{\alpha_1}({\bf r}_1)\cdot ... \cdot
 \zeta_{\alpha_n}({\bf r}_n)}{\int^{'}D\zeta_1 ... \int^{'}D\zeta_n  e^{-\beta[\Omega_{co}[\{\beta J\}]-
\int_{\bf r}J_{\alpha}({\bf r})\zeta_{\alpha}({\bf r})]}}
\\
\nonumber
=\frac{1}{V_S}\int_{{\bf r'}\in S_R({\bf r}_1)}...\frac{1}{V_S}\int_{{\bf r^{(n)}}\in S_R({\bf r}_n)}\langle 
\hat\zeta_{\alpha_1}({\bf r}')\cdot ... \cdot \hat\zeta_{\alpha_n}({\bf r}^{(n)})\rangle,
\end{eqnarray}
with analogous relation for the correlation functions for the microscopic and the mesoscopic densities. 
Note the difference between the micro- and the mesoscopic correlation functions resulting from the integration 
of the former over mesoscopic volumes. In particular, for $2R>\sigma_{\alpha\beta}$ 
the mesoscopic two-point correlation function for densities, analogous to Eq.(\ref{corfue}), does not vanish 
for ${\bf r}_1={\bf r}_2$. This is because for ${\bf r}'\in S_R({\bf r}_1)$ and  
${\bf r}''\in S_R(({\bf r}_1)$, such that $|{\bf r}'-{\bf r}''|>\sigma_{\alpha\beta}$,
 the corresponding microscopic correlation function
 on the RHS of an equation analogous to Eq.(\ref{corfue}) does not vanish and contributes to the integral.
We should stress that in Ref.\cite{ciach:08:1} the theory is based on the mesoscopic density, but for 
significantly different sizes of particles the volume fraction is more convenient, as discussed in the previous subsection.

Let us introduce the Legendre transform
\begin{equation}
 \label{lege}
\beta F[\{\bar\zeta\}]:=\beta\Omega[\{\beta J\}]+\int_{\bf r}\beta J_{\alpha}({\bf r})\bar\zeta_{\alpha}({\bf r})
\end{equation}
where 
\begin{equation}
\label{Phi}
 \bar\zeta_{\alpha}({\bf r})=\frac{\delta(-\beta\Omega)}{\delta(\beta J_{\alpha}({\bf r}))}
\end{equation}
is the  average field  (volume fraction) for given $\{J\}$. The equation of state  takes the form
\begin{equation}
 \label{EOS}
\frac{\delta (\beta F)}{\delta\bar\zeta_{\alpha}({\bf r})}=\beta J_{\alpha}({\bf r}).
\end{equation}

In general $\bar\zeta$ may differ from any mesostate defined in Eq.(\ref{me}).
We extend the functional $\Omega_{co}$ beyond the set of the mesostates. Let the extension be
 defined in Eq.(\ref{Omco}) on the Hilbert space of fields  that fulfill the restrictions following from 
the properties of the mesostates, and let us keep the notation $\Omega_{co}$ for this extension. 
 The key restriction on the mesoscopic volume fraction is the magnitude and the gradient. 
In Fourier representation we shall consider the functions that vanish for $k>\pi/R$, where $k$ is the wave number. 
As discussed in Ref.\cite{ciach:08:1}, the fields with magnitudes exceeding the close packing 
(such fields belong to the Hilbert space, but do not represent any mesostate) are irrelevant, 
since the corresponding Boltzmann factor is very small. We introduce the functional $\beta F[\{\zeta\}]$ of the form
\begin{equation}
\label{FF}
 \beta F[\{\zeta\}]=\beta\Omega_{co}[\{\zeta\}]-\log\Big[\int D\phi_1 ... \int
 D\phi_n e^{-\beta[H_{fluc}-\int_{\bf r} J_{\alpha}({\bf r})\phi_{\alpha}({\bf r})]}\Big],
\end{equation}
where $\phi_{\alpha}({\bf r})$ is the local 
fluctuation of the volume fraction of the component $\alpha$, and
\begin{eqnarray}
\label{Hfl}
H_{fluc}[\{\zeta\},\{\phi\}]=\Omega_{co}[\{\zeta+\phi\}]-\Omega_{co}[\{\zeta\}].
\end{eqnarray}
We introduced the notation  $\{\phi\}=\{\phi_1({\bf r}), ...,\phi_n({\bf r})\}$. From Eqs.(\ref{lege}) and (\ref{OmJ})
 it follows that the functional (\ref{FF})  equals the grand potential,
 when  $\{\zeta\}=\{\bar\zeta\}$, with $\{\bar\zeta\}$ determined from Eq.(\ref{EOS}). 
By definition $\langle \phi_{\alpha}\rangle=0$ when  $\{\zeta\}=\{\bar\zeta\}$.

Note that from Eq.(\ref{FF}) it follows that the inverse correlation functions (related to the direct correlation functions) defined by
 \begin{equation}
\label{calCn}
 C^m_{\alpha_1,...\alpha_n}({\bf r}_1,...,{\bf r}_n)= \frac{\delta^n 
\beta F[\{\bar\zeta\}]}{\delta\bar\zeta_{\alpha_1}({\bf r}_1)...\delta\bar\zeta_{\alpha_n}({\bf r}_n)}
 \end{equation}
consist of two terms: the first one is the contribution from the fluctuations on the microscopic length scale ($<R$) 
with frozen fluctuations on the mesoscopic length scale. This term is
 \begin{equation}
  \label{calCn0}
 {\cal C}^{co}_{\alpha_1,...\alpha_n}({\bf r}_1,...,{\bf r}_n)= \frac{\delta^n
 \beta \Omega_{co}[\{\bar\zeta\}]}{\delta\bar\zeta_{\alpha_1}({\bf r}_1)...\delta\bar\zeta_{\alpha_n}({\bf r}_n)}.
 \end{equation}
 The second term is the contribution from the fluctuations on the mesoscopic length scale ($>R$).
From Eqs. (\ref{OmJ})-(\ref{calCn}) we obtain equations relating the inverse correlation
 functions with the many-body correlation
 functions. In the lowest nontrivial order beyond the mean-field approximation and for $J_{\alpha}=0$ we obtain
 (see \cite{ciach:08:1})
\begin{eqnarray}
\label{avdgg}
 \frac{\delta\beta\Omega_{co}[\{\bar\zeta\}]}{\delta\bar\zeta_{\alpha}({\bf r}) }+
\int_{\bf r_1}\int_{\bf r_2}G^{m}_{\alpha_1\alpha_2}({\bf r}_1,{\bf r}_2)
{\cal C}_{\alpha_1\alpha_2}^{co}({\bf r}_1,{\bf r}_2,{\bf r})=0,
\end{eqnarray}
and
\begin{eqnarray}
\label{CalC2}
 2C^m_{\alpha\beta}({\bf r}_1,{\bf r}_2)= {\cal C}_{\alpha\beta}^{co}({\bf r}_1,{\bf r}_2)+
\langle \frac{\delta^2(\beta H_{fluc})}{\delta\zeta_{\alpha}({\bf r}_1)\delta\zeta_{\beta}({\bf r}_2)}\rangle 
-\langle \frac{\delta(\beta H_{fluc})}{\delta\zeta_{\alpha}({\bf r}_1)}\frac{\delta(\beta H_{fluc})}
{\delta\zeta_{\beta}({\bf r}_2)}\rangle^{con}+\\
\nonumber
\int_{\bf r'}\Bigg[\langle \frac{\delta H_{fluc}}{\delta\zeta_{\alpha}({\bf r}_1)}\phi_{\alpha_1}({\bf r}')
\rangle C^m_{\alpha_1 \beta}({\bf r}',{\bf r}_2)+
\langle \frac{\delta H_{fluc}}{\delta\zeta_{\beta}({\bf r}_2)}\phi_{\beta_1}({\bf r}')
\rangle C^m_{\beta_1\alpha}({\bf r}',{\bf r}_1)
\Bigg].
\end{eqnarray}

Eq. (\ref{avdgg}) is the minimum condition for the grand potential. In the MF approximation the second term in Eq.(\ref{avdgg}) is 
neglected. Since there may exist several local minima, the solution corresponds to a stable or to a metastable
 phase when the grand potential assumes the global or the local minimum respectively. The solution of Eq.(\ref{avdgg}) 
corresponding to the global minimum gives the average density for given $\mu$ and $T$ in the lowest nontrivial order 
beyond MF.

In order to obtain the two-point inverse correlation function from Eq.(\ref{CalC2}), 
 $H_{fluc}[\{\zeta\},\{\phi\}]$ in Eq.(\ref{Hfl}) is expanded in $\phi_{\alpha}$, and the expansion is truncated.
 Since the volume fractions are less than unity, the corresponding fluctuations are small and
 such an expansion is justified.
In this way an equation relating the two-point  inverse correlation function with many-body correlation functions is obtained.
 Approximate equation that can be solved in practice will be derived in the next section. 
From Eqs.(\ref{EOS}),(\ref{Phi}) and (\ref{genfu}) we obtain the analog of the Ornstein-Zernike equation
\begin{equation}
 \label{OZ1}
\int_{\bf r_2}C^m_{\alpha\alpha_1}({\bf r}_1,{\bf r}_2)G^{m}_{\alpha_1\beta}({\bf r}_2,{\bf r}_3)
=\delta({\bf r}_1-{\bf r}_3)\delta^{Kr}_{\alpha\beta}.
\end{equation}

\subsection{Periodic structures} 
Let us consider  periodic density profiles
\begin{eqnarray}
\label{perdens}
 \bar\zeta_{\alpha}({\bf r})=\bar\zeta_{\alpha}^0+\Phi_{\alpha}({\bf r})
\end{eqnarray}
where
\begin{eqnarray}
 \Phi_{\alpha}({\bf r}+{\bf P})=\Phi_{\alpha}({\bf r})
\end{eqnarray}
and ${\bf P}=\sum_{i}^3n_i{\bf p}_i$ where ${\bf p}_i$ are the vectors connecting the centers of the nearest-neighbor 
unit cells and  $n_i$ are integer numbers. The $\bar\zeta^0_{\alpha}$ is the space-averaged density, i.e.
\begin{eqnarray}
\label{Pp}
 \int_{{\bf r}\in {{\cal V}_u}}\Phi_{\alpha}({\bf r})=0,
\end{eqnarray}
where ${\cal V}_u$ is the unit cell of the periodic structure, whose volume is denoted by $V_u$.
In the case of  periodic structures 
\begin{eqnarray}
\label{CPe}
C^m_{\alpha\beta}({\bf r}_1+{\bf P},{\bf r}_2+{\bf P})=C^m_{\alpha\beta}({\bf r}_1,{\bf r}_2)
=C^m_{\alpha\beta}(\Delta{\bf r}|{\bf r}_2)
\end{eqnarray}
 where 
$\Delta{\bf r}={\bf r}_1-{\bf r}_2\in R^3$ and ${\bf r}_2\in {\cal V}_u$.  We introduce the inverse correlation
 function averaged over the unit cell by 
\begin{eqnarray}
\label{CCga1}
 C_{\alpha\beta}(\Delta{\bf r})=\frac{1}{V_u}\int_{{\bf r_2}\in {\cal V}_u}
C^{m}_{\alpha\beta}(\Delta{\bf r}|{\bf r}_2)
\end{eqnarray}
with analogous definition for the correlation function $ G_{\alpha\beta}$ averaged over the unit cell (in terms of
$ G^{m}_{\alpha\beta}$). In Fourier representation
 from Eq.(\ref{OZ1}) we have \cite{ciach:08:1}
\begin{equation}
\label{OZF}
 \tilde C_{\alpha\gamma}({\bf k})\tilde G_{\gamma\beta}({\bf k})=\delta^{Kr}_{\alpha\beta}.
\end{equation}
We decompose $H_{fluc}$ into two parts,
\begin{equation}
\label{Har}
H_{fluc}[\{\bar\zeta\},\{\phi\}]
={\cal H}_{G}[\{\bar\zeta\},\{\phi\}]+\Delta{\cal H}[\{\bar\zeta\},\{\phi\}].
\end{equation}
The first term in the above equation is given by
\begin{eqnarray}
\label{Haga}
 {\cal H}_G[\bar\zeta,\phi]=
\frac{1}{2}\int_{\bf k}\tilde \phi_{\alpha}({\bf k})\tilde C_{\alpha\beta}({\bf k})\tilde\phi_{\beta}(-{\bf k})
=\frac{1}{2}\int_{\bf k}\tilde \psi_i({\bf k})\tilde C_i({\bf k})\tilde \psi_i(-{\bf k}),
\end{eqnarray}
where $\tilde C_i({\bf k})$  and $\tilde \psi_i({\bf k})$ are the eigenvalues and the eigenmodes respectively of 
the  matrix $\tilde{\bf C}$ with the elements $\tilde C_{\alpha\beta}({\bf k})$,
and summation convention for $i $ is used. For brevity we introduced the notation
$ \int_{\bf k}\equiv \int\frac{d{\bf k}}{(2\pi)^3}$. In the next step we make an assumption that $\Delta{\cal
 H}[\bar\zeta,\phi]$ can be treated as a small perturbation. When such an
 assumption is valid, we obtain \cite{podneks:96:0,ciach:06:2}
\begin{eqnarray}
\label{OmHarval}
\beta\Omega[\bar\zeta]\approx \beta\Omega_{co}[\bar\zeta]-\log\int D\phi_1... \int D\phi_n
e^{-\beta{\cal H}_G}
+\langle \beta\Delta{\cal H}\rangle_G
+O(\langle \beta\Delta{\cal H}\rangle_G^2).
\end{eqnarray}
where $\langle ...\rangle_G$ denotes averaging with the Gaussian Boltzmann
factor $e^{-\beta{\cal H}_G}$. 
 Eqs. (\ref{CalC2}) - (\ref{OmHarval})
 allow for calculation of the fluctuation contribution to the grand potential in the lowest nontrivial order when
 the form of $\Omega_{co}$ is known, and  the form of $\tilde C_{\alpha\beta}$ (Eqs.(\ref{CCga1}) and (\ref{calCn})) 
is determined by a self-consistent solution of some approximate version of Eq.(\ref{CalC2}). 
In general, each contribution to Eq.(\ref{OmHarval}) depends on the mesoscopic length scale $R$, but the $R$-dependent 
contributions must cancel against each other to yield  $R$-independent $\Omega$. 
By minimizing the density functional (\ref{OmHarval}) we find the equilibrium structure.
 The main difficulty consists in determination of $\tilde C_i({\bf k})$. 

\section{Approximate theory for weak ordering}
\subsection{ Grand potential and  correlation functions in the case of weak ordering}

If ordering in the system occurs on a length scale larger than the size of particles,
 the local density approximation can be applied, and we assume
\begin{eqnarray}
\label{Fhfh}
 F_h[\{\zeta\}]=\int_{\bf r}f^h(\zeta_1({\bf r}),...,\zeta_n({\bf r})),
\end{eqnarray}
where $f^h(\zeta_1({\bf r}),...,\zeta_n({\bf r}))$ is the free-energy density of the hard-sphere system in which the 
volume fractions in 
the infinitesimal volume $d{\bf r}$ at ${\bf r}$ are $\zeta_1({\bf r}),...,\zeta_n({\bf r})$ . 

In this approximation  we obtain the  functionals
\begin{eqnarray}
\label{Omcoap}
 \beta\Omega_{co}[\{\zeta\}]=\frac{1}{2}\int_{{\bf r}_1}\int_{{\bf r}_2}\beta  V_{\alpha\beta}^{co}(r_{12})
\zeta_{\alpha}({\bf r}_1)\zeta_{\beta}({\bf r}_2)-\int_{\bf r}\beta f_h(\{\zeta\})-
\int_{\bf r}\beta\bar\mu_{\alpha}\zeta_{\alpha}({\bf r}),
\end{eqnarray}
and
\begin{eqnarray}
\label{DelOM}
 \beta H_{fluc}[\{\bar\zeta\},\phi]=
\int_{\bf r_1}\beta\Big[f^h_{\alpha_1}(\bar\zeta({\bf r}_1))-\bar\mu_{\alpha_1} +\int_{{\bf r}_2}
 V_{\alpha_1\alpha_2}^{co}(r_{12})\bar\zeta_{\alpha_2}({\bf r}_2)\Big] 
\phi_{\alpha_1}({\bf r}_1)
\\
\nonumber
+\frac{1}{2}\int_{\bf r_1}\int_{\bf r_2}\phi_{\alpha}({\bf r}_1) {\cal C}^{co}_{\alpha\beta}({\bf r}_1,{\bf r}_2)
\phi_{\beta}({\bf r}_2)+
\sum_{n=3}\int_{\bf r}\frac{\beta f^h_{\alpha_1,...,\alpha_n}(\bar\zeta({\bf r}))}{n!}\phi_{\alpha_1}({\bf r})... 
\phi_{\alpha_n}({\bf r}).
\end{eqnarray}
where $ V_{\alpha_1\alpha_2}^{co}(r_{12})$ is defined in Eq.(\ref{U}) and
\begin{eqnarray}
\beta f^h_{\alpha_1,...,\alpha_n} (\{\zeta({\bf r})\})=
\frac{\partial^n \beta f^h(\{\zeta({\bf r})\})}{\partial \zeta_{\alpha_1}({\bf r})...\partial \zeta_{\alpha_n}({\bf r})}
\end{eqnarray}
 depends on the (local) composition of the mixture, but is independent of temperature.
 For inhomogeneous phases in soft matter systems (colloidal crystals for example), with position-dependent volume 
fractions,  determination of the grand potential and the average distribution of particles within this theory is 
still very difficult. However, further simplifying assumptions can be made in the case of weak ordering. 
 In the case of 'soft' crystalline phases with unit cells of the structure significantly larger than the size of
 the particles, particles (and the whole clusters) can fluctuate around their average positions. The displacements of the particles from their
 average positions can be large, but the long-range order can be preserved. Averaging over such fluctuations leads to 
smooth functions $\Phi_{\alpha}({\bf r})$ with small magnitudes
$\Phi_{\alpha}\ll\zeta_{\alpha}^0 $ (see Eq.(\ref{perdens})).
 In the case of 'soft' crystalline phases the functional (\ref{OmHarval}) can be approximated by
\begin{eqnarray}
\label{OmHarval1}
\beta\Omega[\{\zeta^0+\Phi({\bf r})\}]/V\approx \beta\Omega_{co}[\{\zeta^0+\Phi({\bf r})\}]/V\\
\nonumber
+\frac{1}{2}\int_{\bf k}\sum_{i=1}^n\Bigg[\ln\Bigg(\frac{\tilde C_i(k)}{2\pi}\Bigg)
+
\tilde C_i^{co}(k)\tilde G_i( k)-1\Bigg]\\
\nonumber
+\frac{ \beta f^h_{\alpha_1\alpha_2\alpha_3\alpha_4}(\{\bar\zeta^0\})
{\cal G}^{\alpha_1\alpha_2}{\cal G}^{\alpha_3\alpha_4}}{8},
\end{eqnarray}
where   $\tilde C_i^{co}(k)$ and $\tilde G_i( k)$ denote the eigenvalues of the matrices $\tilde {\bf C}^{co}(k)$
 and $\tilde {\bf G}(k)=\tilde {\bf C}^{-1}(k)$ respectively, with the element $(\alpha,\beta)$ of the former given by
\begin{eqnarray}
\label{Cco2g}
 \tilde C_{\alpha\beta}^{co}(k)=\tilde  C_{\alpha\beta}^{0}(k)+
\frac{\beta f^h_{\alpha\beta\alpha_1\alpha_2}(\bar\zeta^0)}{2}\int_{{\bf r}\in{\cal V}_u}\frac{\Phi_{\alpha_1}({\bf r})
\Phi_{\alpha_2}({\bf r})}{V_u},
\end{eqnarray}
where $\zeta^0_{\alpha}+\Phi_{\alpha}({\bf r})$ characterizes
local volume fraction of the $\alpha$-th component in the inhomogeneous ordered phase, and
\begin{eqnarray}
\label{Cco2g1}
 \tilde C_{\alpha\beta}^{0}(k)=\beta\tilde V_{\alpha\beta}^{co}(k)+
\beta f^h_{\alpha\beta}(\{\bar\zeta^0\}).
\end{eqnarray}
In the disordered phase $\Phi_{\alpha}({\bf r})=0$ for each component $\alpha$, and $ \tilde C_{\alpha\beta}^{co}(k)$ 
reduces to $ \tilde C_{\alpha\beta}^{0}(k)$.
Finally,
\begin{eqnarray}
\label{calG}
 {\cal G}^{\alpha\beta}=\int_{\bf k}\tilde G_{\alpha\beta}(k).
\end{eqnarray}
 Recall that by construction of the mesoscopic theory on the length scale $R$, 
the cutoff $\sim\pi/R$ 
is present in the  integral in Eq.(\ref{calG}). Recall also that from Eq.(\ref{corfue}) and discussion in sec.2.1
 it follows that
$\int_{\bf k}\tilde G_{\alpha\beta}(k)=G_{\alpha\beta}(0)$ \textit{differs from the microscopic correlation function
 at zero distance} and is finite. Since the mesoscopic length scale is larger than the size of  molecules and 
 smaller than the length scale of the ordering  but otherwise it is arbitrary, 
the above approximate version of the theory is valid as long as the $R$-dependent terms in Eq.(\ref{calG}) are negligible
 compared to the dominant contribution. 

When $\Phi_{\alpha}\ll\zeta_{\alpha}^0 $, Eq.(\ref{Omcoap}) can be approximated by the expression
\begin{eqnarray}
\label{OmcoBr}
 \beta\Omega_{co}[\{\zeta^0+\Phi({\bf r})\}]=\beta\Omega_{co}[\{\zeta^0\}] +\beta\Omega_G[\{\zeta^0+\Phi({\bf r})\}]\\
\nonumber
+
\sum_{n\ge 3}\frac{ \beta f^h_{\alpha_1,...,\alpha_n}(\{\zeta^0\})}{n!}
\int_{\bf r}\Phi_{\alpha_1}({\bf r})...\Phi_{\alpha_n}({\bf r})
\end{eqnarray}
with
\begin{eqnarray}
 \label{OmcoBrG}
 \beta\Omega_G[\{\zeta^0+\Phi({\bf r})\}]=
\frac{1}{2}\int_{\bf k}\tilde\Phi_{\alpha}({\bf k})\tilde C_{\alpha\beta}^{0}(k)\tilde\Phi_{\beta}(-{\bf k})=
\frac{1}{2}\int_{\bf k}\tilde\Psi_i({\bf k})\tilde C_i^{0}(k)\tilde\Psi_i(-{\bf k}),
\end{eqnarray}
where $\tilde C_i^{0}(k)$ and $\tilde\Psi_i({\bf k})$ are the eigenvalues and eigenvectors of the matrix 
$\tilde{\bf C}^0$ respectively.

In order to obtain approximation for $\tilde C_{\alpha\beta}$ from Eq.(\ref{CalC2}), we  truncate the expansion of
 $H_{fluc}[\{\zeta\},\{\phi\}]$ (see Eqs.(\ref{Hfl}) and (\ref{DelOM})) at the term $O(\phi^4)$.
 Next, the four- and six-point correlation
 functions are approximated by products of two-point correlation functions, and after some  algebra 
we obtain the approximate result, valid for periodic structures 
\begin{eqnarray}
\label{CP}
2\tilde{\bf C}(k)=(\tilde{\bf C}^{co}(k)+{\bf A})\big[3{\bf I}-\tilde{\bf G}(k)(\tilde{\bf C}^{co}(k)+{\bf A})\big]-
\tilde{\bf P}(k)
\end{eqnarray}
where ${\bf I}$ is the unitary matrix ($I_{\alpha\beta}=\delta^{Kr}_{\alpha\beta}$), and
 the $(\alpha,\beta)$ element of the matrix ${\bf A}$ is 
\begin{eqnarray}
\label{A}
 A_{\alpha\beta}=
\frac{1}{2}\beta f^h_{\alpha\beta\gamma\nu}(\{\bar\zeta^0\})\int_{\bf k}\tilde G_{\gamma\nu}(k).
\end{eqnarray}
Note that ${\bf A}$ is independent of $k$. Finally, 
 the $(\alpha,\beta)$ element of the matrix ${\bf P}$ is
\begin{eqnarray}
  \tilde P_{\alpha\beta}(k)=\frac{\beta f^h_{\alpha\alpha_1\alpha_2}\beta f^h_{\beta\beta_1\beta_2}}{2}\int_{\bf r}
e^{i{\bf k}\cdot{\bf r}}G_{\alpha_1\beta_1}(r)G_{\alpha_2\beta_2}(r)\\
\nonumber
+\frac{\beta f^h_{\alpha\alpha_1\alpha_2\alpha_3}\beta f^h_{\beta\beta_1\beta_2\beta_3}}{6}
\int_{\bf r}e^{i{\bf k}\cdot{\bf r}}G_{\alpha_1\beta_1}(r)G_{\alpha_2\beta_2}(r)G_{\alpha_3\beta_3}(r).
\end{eqnarray}
 When ${\bf P}$ is neglected in Eq.(\ref{CP}), then we obtain a simple equation for
 $\tilde {\bf C}(k)$,
 analogous to the self-consistent Hartree approximation and Brazovskii theory \cite{brazovskii:75:0} 
generalized for mixtures (linear 
approximation with respect to $\beta f^h_{\alpha_1...\alpha_n}$),
\begin{eqnarray}
\label{CF}
\tilde {\bf C}(k)=\tilde {\bf C}^{co}(k)+ {\bf A}.
\end{eqnarray}
By inserting Eq.(\ref{CF}) into Eq.(\ref{CP}) one can easily check  validity of the former when 
${\bf P}$ is neglected. Note that the dependence on $k$ in Eq.(\ref{CF}) is not changed compared to the MF result, 
because ${\bf A}$ is independent of $k$. However, when ${\bf P}$ is included in Eq.(\ref{CP}),
 the dependence on $k$ is different
 than in MF. In Ref.\cite{ciach:08:1} the expansion of $H_{fluc}$ was truncated at the 
term $O(\phi^2)$, which leads to less accurate approximation. However, at linear order in 
derivatives of $f^h$ the same self-consistent Hartree approximation
 (Eq.(\ref{CF})) was obtained and used in further applications.
\subsection{Boundary of stability of the homogeneous phase in the Brazovskii-type approximation}

The homogeneous system is unstable with respect to an infinitesimal fluctuation $\tilde\psi_i(k)$
 when $\tilde C_i(k)<0$. The instability occurs when the  fluctuation $\tilde\psi_i(k)$ is
 excited and at the same time  $\tilde \psi_j(k)=0$ for $j\ne i$.  The boundary of stability of the 
homogeneous phase is given by 
\begin{eqnarray}
\label{detC}
\det \tilde {\bf C}(k_b)=\prod_{i=1}^n\tilde C_i(k_b) =0,
\end{eqnarray}
 where $ k_b$ corresponds to the highest temperature for which any instability occurs for given 
composition of the mixture $\{\bar \zeta^0\}$. Since the temperature 
at the instability with respect to the concentration wave $\tilde\psi_i(k)$ with the wave-number $k$, $T(k)$,  
is given by $ \det \tilde {\bf C}(k)=0$, the (local)  maximum condition $dT/dk=0$ is equivalent to 
\begin{eqnarray}
\label{detCder}
\frac{ d\det \tilde {\bf C}(k)}{dk}|_{k=k_b}=0.
\end{eqnarray}
If there are several solutions of Eqs. (\ref{detC}) and (\ref{detCder}), the one corresponding to the highest 
temperature for given composition $\{\bar \zeta^0\}$ determines the temperature and the
wave number of the critical mode at the boundary of stability of the homogeneous phase with respect to concentration 
fluctuations with infinitesimal amplitudes.

In MF Eq.(\ref{detC}) reduces to $\det \tilde {\bf C}^0(k_b)=0$, which  for fixed composition  $\{\bar \zeta^0\}$ 
 in an $n$-component system is an $n$-th order equation for $\beta=1/(k_BT)$  (see Eq.(\ref{Cco2g1}) and note 
that $\beta f^h_{\alpha_1\alpha_2}$ and $ \tilde V_{\alpha\beta}^{co}(k)$ are independent of $T$). 
There are up to $n$ solutions for $T$ for fixed composition and $k$, depending on the interaction potentials.
In the one-component system 
there is one such solution, and if it corresponds to $k_b>0$, the corresponding line $T(\zeta)$
is known as the $\lambda$-line \cite{ciach:00:0,ciach:03:1,ciach:08:1}. In multicomponent system we should speak about 
$\lambda$-surface.

 In the mesoscopic theory the probability of a deviation from the average
 composition  $\{\bar \zeta^0\}$ on the  mesoscopic length scale, $\{\phi({\bf r})\}$, is proportional to
 $\exp(-\beta\Omega_{co}[\{\phi({\bf r})+\bar \zeta^0\}])$. Inhomogeneous distribution of the particles
 on the mesoscopic length scale can be more probable than the homogeneous states when $\beta\Omega_{co}$ 
does not assume a minimum for $\{\phi({\bf r})=0\}$, i.e. when  $\det \tilde {\bf C}^0(k_b)<0$. As discussed
 in Refs.\cite{ciach:00:0,ciach:08:1,ciach:06:1}, at the  $\lambda$-line (or $\lambda$-surface) a change from locally
 homogeneous to locally periodic structure occurs, because for  $\det \tilde {\bf C}^0(k_b)<0$ the waves with
 the wavelength $2\pi /k_b$ (and infinitesimal amplitudes) are more probable than the constant volume fractions. 
However, in the presence of mesoscopic fluctuations the averaging over different waves of concentration may lead 
to position-independent {\it average} volume fractions, by which  the stability of the disordered phase can be restored. 

The open question for a multicomponent system is whether solutions of 
Eqs.(\ref{detC}) and (\ref{detCder})
 exist beyond MF. To answer this question let us focus on  $\tilde {\bf C}$ ( Eqs.(\ref{CF}),
 (\ref{Cco2g})  and (\ref{A})), and note that the matrix
${\bf A}$ is a linear combination of the integrals of the 
 correlation functions,
\begin{eqnarray}
\tilde G_{\alpha\beta}(k) \propto \frac{1}{\det \tilde {\bf C}(k)}.
\end{eqnarray}
 $\det \tilde {\bf C}(k)$  can be Taylor expanded near the minimum at   $k_b$, 
\begin{eqnarray}
 \det \tilde {\bf C}(k)\approx \det \tilde {\bf C}(k_b)+c_2(k-k_b)^2+O((k-k_b)^3)
\end{eqnarray}
 and from Refs.\cite{brazovskii:75:0,patsahan:07:0,ciach:08:1} we obtain 
\begin{eqnarray}
\int_{\bf k}\tilde G_{\gamma\nu}(k)\propto
\left\{ 
\begin{array}{lll}
  \pi/R &\;\; {\rm for} &\;\;k_b=0\\  
\frac{ k_b^2}{\sqrt {\det \tilde {\bf C}(k_b)}}+O(1/R) &\;\; {\rm for} &\;\;k_b>0
\end{array}
\right.
 \label{CoPY}.
\end{eqnarray}
  The cases  $k_b=0$ and $k_b>0$, 
corresponding to macro- and micro-phase separation respectively, are
 significantly different. 
In the case of separation into two homogeneous phases the temperature at the instability is shifted compared to the MF result, 
because the matrix elements of ${\bf A}$ are finite. Since $ A_{\alpha\beta}\propto \pi/R$, the shift
 depends on the scale of coarse-graining. This approximation is oversimplified for precise determination 
of the spinodal line for the macroscopic phase separation. We can determine the upper bound of the shift,
 because $R\ge \sigma_{\alpha\beta}$.
When the interaction potentials are such that Eq.(\ref{detCder})
 leads to $k_b>0$, then from Eq.(\ref{CoPY}) it follows that the matrix ${\bf A}$ becomes
 singular if $\det \tilde {\bf C}(k_b)\to 0$,  and
 Eq. (\ref{detC}) has no solutions for finite $\tilde {\bf C}^{0}(k_b)$. Only for $T\to 0$,
 i.e. when $\tilde {\bf C}^{0}(k_b)$ becomes singular as well, the solution of  Eq.(\ref{detC}) may exist.

At first order transitions between inhomogeneous phases the grand-potential density takes
 the same value for different phases. Inhomogeneous phases may appear when the periodic mesoscopic
 densities are more probable than the uniform states, i.e. in the phase-space 
region where $\det \tilde {\bf C}^{0}(k_b)<0$. For such thermodynamic states the fluctuations on the
 mesoscopic length scale dominate, because the corresponding contribution to $ \tilde {\bf C}$ is large enough to yield 
 $\det \tilde {\bf C}>0$ despite $ \det\tilde {\bf C}^{0}<0$. Thus, in order to obtain the phase diagram precisely,
 the fluctuation contribution to the grand potential cannot be neglected.
 We should note that the relatively simple MF 
stability analysis allows for the rough estimation of the phase space part where the structure may be inhomogeneous on the mesoscopic length scale.

\section{Two-component mixture}
The general results of the previous sections apply in particular to two-component mixtures. Two component mixtures were studied in Ref.\cite{patsahan:99:0} within the method of collective variables \cite{yukhnovskii:78:0} under the assumption of hmomgeneous structure. Here we ara mainly interested in inhomogeneous fluids.
In the first step we need to determine the inverse correlation functions in MF approximation, 
and this is a subject of this section.
The form of the free energy for hard spheres in two-component systems is known \cite{lebowitz:64:0}, 
and was used recently in the case of ionic systems with size asymmetry of ions \cite{ciach:07:0,patsahan:10:0}. 
Hence, we can perform more detailed analysis within the framework developed above,
 still for arbitrary form of the interaction potentials.

It is convenient to introduce
\begin{equation}
 \sigma_{\alpha\beta}=\frac{\sigma_{\alpha}+\sigma_{\beta}}{2}.
\end{equation}
As a length unit we choose $\sigma_{12}$, and the wave-numbers are in $\sigma_{12}^{-1}$ units. 
We shall use the index $\alpha= 1 $ for the larger, and the index  $\alpha= 2$ for the smaller particle.
The asymmetry of the size of the particles can be characterized by 
\begin{equation}
\label{rpm}
r_{1}=\frac{\sigma_{1}}{\sigma_{12}}=1+ \delta \hskip1cm r_{2}=\frac{\sigma_{2}}{\sigma_{12}}=1- \delta,
\end{equation}
where $0\le \delta\le 1$; $\delta=0,1$ for identical sizes of the particles and for point-like smaller particles
 respectively. We also introduce volume fraction of both components,
\begin{equation}
\zeta=\zeta_1+\zeta_2 .
\end{equation}
We shall consider dimensionless correlation functions for 
 local deviations of the volume fractions from the space-averaged values, 
$\Phi_{\alpha}({\bf r})=\zeta_{\alpha}({\bf r})-\zeta^0_{\alpha}$.
\subsection{Mean field approximation for arbitrary interaction potentials}
The partial inverse correlation functions in the disordered phase in MF approximation, $\tilde C_{\alpha\beta}^0(k)$, 
 are given in Eq.(\ref{Cco2g}). Explicit expressions for dimensionless partial derivatives of the
free-energy density for hard-sphere reference system,
 $  f^*_{\alpha\beta}(\zeta_1,\zeta_2)=\sigma_{12}^3 \beta f^h_{\alpha\beta}(\zeta_1,\zeta_2)$,
 are given in Appendix. In the case of ordering on the
mesoscopic length scale we make the approximation (\ref{VU}) for the interaction potentials in Eq.(\ref{Vco}).
 When the microscopic structure is disregarded, 
the microscopic correlation function takes the form 
\begin{eqnarray}
\label{gtheta}
  g_{\alpha\beta}^{co}({\bf r}_1-{\bf r}_2)=\theta(|{\bf r}_1-{\bf r}_2|-\sigma_{\alpha\beta}),
\end{eqnarray}
and we approximate the interaction-potential density by
\begin{eqnarray}
\label{VU1}
  V^{*}_{\alpha \beta}({\bf r},{\bf r}') =
\sigma_{12}^6 V^{co}_{\alpha \beta}({\bf r},{\bf r}')
\approx\Bigg(\frac{6}{\pi}\Bigg)^2\frac{U_{\alpha \beta}({\bf r},{\bf r}')\theta(|{\bf r}_1-{\bf r}_2|
-\sigma_{\alpha\beta})}{r^3_{\alpha}r^3_{\beta}}.
\end{eqnarray}

Eqs.(\ref{detC}) and (\ref{detCder}), from which the temperature at the boundary of stability of the
 disordered phase can be obtained, for the two-component mixtures take the forms
\begin{equation}
\label{i1}
 \beta^2 \tilde U(k)+\beta\tilde  K(k)+D=0
\end{equation}
and 
\begin{equation}
\label{i2}
  \beta\tilde U'(k)+ \tilde K'(k)=0,
\end{equation}
where the prime denotes a derivative with respect to $k$, and we have introduced 
\begin{equation}
 D=\det {\bf F}\hskip1cm \tilde U(k)=\det \tilde {\bf V}^{*}(k)
\end{equation}
 and
\begin{equation}
\tilde K(k)=\tilde V_{11} ^{*}(k) f^*_{22}+\tilde V ^{*}_{22}(k) f^*_{11}
-2\tilde V ^{*}_{12}(k) f^*_{12}=
D Tr (\tilde {\bf V} ^{*}(k) {\bf F}^{-1}).
\end{equation}
 By $\tilde {\bf V}^{*}(k)$ and ${\bf F}$ we denote the matrices with the  $(\alpha,\beta)$
 element given by $\tilde V_{\alpha\beta}^{*}(k)$ and $ f^*_{\alpha\beta}$ respectively.
 $D$ can be directly calculated from ${\bf F}$ given in Appendix, and for any size ratio has the simple form
\begin{equation}
 D=\frac{(1+2\zeta)^2}{\zeta_1\zeta_2(1-\zeta)^4}>0.
\end{equation}

Let us discuss conditions under which solutions of Eq.(\ref{i1}) for $1/\beta$,
\begin{equation}
 k_BT_{1,2}=\frac{2\tilde U(k)}{-\tilde K(k)
\pm \sqrt{\tilde K(k)^2-4D\tilde U(k)}},
\end{equation}

 are real  positive numbers. 
Two types of interaction potentials can be distinguished: (i)  $ \tilde U(k)<0$ and (ii)  $ \tilde U(k)>0$.

(i)  Interaction 
potentials such that $ \tilde U(k)<0$  characterize, in particular,  the Primitive Model of ionic 
systems (hard sphere and Coulomb potential) for $k>0$   \cite{ciach:07:0}. Since $D>0$, there is always one positive
solution of Eq.(\ref{i1}), namely $k_BT_2$. 

(ii)  The  case of $ \tilde U(0)>0$ corresponds,
 in particular, to stronger attraction (first moment of the interaction potential)  between like particles than 
between particles of different kinds. For $ \tilde U(k)>0$ the necessary condition for an instability with respect to a density wave with the wavelength
 $k$ is $\tilde  K(k)<0$, because for $\tilde  K(k)>0$ the solutions of Eq.(\ref{i1}) for $\beta$, if exist, are negative. 
Another condition that must be satisfied for existence of solutions of Eq.(\ref{i1}) is $\tilde K(k)^2-4D\tilde U(k)>0$. 
The higher temperature is for the solution  $k_BT_2$.

In order to determine what kind of inhomogeneities appear in the system beyond the boundary of stability 
of the homogeneous phase we focus on
the eigenvalues and eigenvectors of $\tilde{\bf C}^{0}(k)$. The  eigenvalues  depend on the wave-number $k$ 
in a nontrivial way,
\begin{equation}
\label{Cphi}
\tilde C_{1}^0(k)=
\frac{\tilde C^0_{11}(k)+\tilde C^0_{22}(k)
-sign(\tilde C^0_{12}(k))  B(k)}{2}
\end{equation}
and
\begin{equation}
\label{Ceta}
\tilde C_{2}^0(k)=
\frac{\tilde C^0_{11}(k)+\tilde C^0_{22}(k)
+sign(\tilde C^0_{12}(k))  B(k)}{2},
\end{equation}
with
\begin{equation}
\label{calB}
B(k)=\sqrt{ A^2+4\tilde C^0_{12}(k)^2},
\end{equation}
and
\begin{equation}
\label{calA}
A(k)=sign(\tilde C^0_{12}(k))
\big[\tilde C^0_{22}(k)-\tilde C^0_{11}(k)\big].
\end{equation}

The corresponding eigenmodes have the forms
\begin{equation}
\label{v2}
\tilde\Psi_1({\bf k})=\tilde a(k)\tilde \Phi_{1}({\bf k})
-\tilde b(k)\tilde \Phi_{2}({\bf k}),
\end{equation}
\begin{equation}
\label{v1}
\tilde\Psi_2({\bf k})=\tilde b(k)\tilde \Phi_{1}({\bf k})
+\tilde a(k)\Phi_{2}({\bf k}),
\end{equation}
where 
\begin{equation}
\label{a}
\tilde a(k)=\Bigg[\frac{ A(k)+
 B(k)}{2 B(k)}\Bigg]^{1/2},
\end{equation}
\begin{equation}
\label{b}
\tilde b(k)=\Bigg[\frac{- A(k)
+B(k)}{2B(k)}\Bigg]^{1/2}.
\end{equation}
  The above expressions differ from the corresponding expressions
derived for the primitive model of ionic systems in Ref.\cite{ciach:07:0}, because here we consider volume 
fractions rather than number densities. Note that as the fluctuating field either the local number density or
 the local volume fraction can be chosen, and
 the physical properties like phase transitions should be independent of this choice. By changing 
$\zeta_{\alpha} \to \rho_{\alpha}$ we should simultaneously rescale  
$\tilde C_{\alpha\beta}(k)\to \tilde C_{\alpha\beta}(k)v_{\alpha}v_{\beta}$, 
and the Eqs. (\ref{detC}) and (\ref{detCder}) for the rescaled functions yield the same solutions.
 
The eigenmodes represent two order-parameter (OP) fields, and in
 principle either one of them may lead to instability of the
 disordered phase for given $(\zeta_1,\zeta_2)$. For small-amplitude inhomogeneities the last term in 
 Eq.(\ref{OmcoBr}) can be neglected, and we can limit ourselves to the Gaussian approximation in
 Eq.(\ref{OmcoBrG}). The OP  $\tilde \Psi_{i}({\bf k}_b)$ induces the instability 
when $\tilde C_{i}^0(k_b)=0$. Recall that at the boundary of stability with respect to
the fluctuation $\tilde \Psi_{i}({\bf k}_b)$ the other OP vanishes (it would yield a positive contribution to the 
grand potential). 
The requirement that $\tilde \Psi_{j}({\bf k}_b)=0$ for $j\ne i$ leads to the relation between the critical amplitudes (see Eqs.(\ref{v1}) and (\ref{v2}))
\begin{eqnarray}
 \frac{ \tilde \Phi_2({\bf k}_b)}{\tilde \Phi_1({\bf k}_b)} =\left\{ 
\begin{array}{lll}
 -\frac{\tilde b(k_b)}{\tilde a(k_b)
}=-\frac{B(k_b)-A(k_b)}{2\tilde C^0_{12}(k_b)}&\;\; {\rm for} &\;\;\tilde C_{1}^0(k_b)=0
\\ 
\\
\frac{\tilde a(k_b)}{\tilde b(k_b)}
=\frac{B(k_b)+A(k_b)}{2\tilde C^0_{12}(k_b)}.
 &\;\; {\rm for} &\;\;\tilde C_{2}^0(k_b)=0
\end{array}
 \label{r1}.
\right.
\end{eqnarray}
Since $B(k)\pm A(k)>0$, we have 
\begin{eqnarray}
 sign(\tilde \Phi_2({\bf k}_b)\tilde \Phi_1({\bf k}_b))=
\left\{ 
\begin{array}{lll}
 - sign(\tilde C^0_{12}(k_b)) &\;\; {\rm for} &\;\;\tilde C_{1}^0(k_b)=0
\\ 
\\ 
sign(\tilde C^0_{12}(k_b)) &\;\; {\rm for} &\;\;\tilde C_{2}^0(k_b)=0
\end{array}
 \label{ratio}.
\right.
\end{eqnarray}

Let us consider planar waves 
$\Phi_i({\bf r})=\tilde \Phi_i({\bf k}_b)\cos({\bf k}_b\cdot {\bf r})$. The local change of the volume fraction is
 $\zeta({\bf r})-\zeta^0=\tilde \Phi_1({\bf k}_b)\cos({\bf k}_b\cdot {\bf r})+
\tilde \Phi_2({\bf k}_b)\cos({\bf k}_b\cdot {\bf r})$. When $\tilde \Phi_2({\bf k}_b)\tilde \Phi_1({\bf k}_b)>0$, 
then dense regions of size $\pi/k_b$ where
 $\zeta({\bf r})-\zeta^0>0$  are followed by dilute regions of similar size where $\zeta({\bf r})-\zeta^0<0$.  
When $\tilde \Phi_2({\bf k}_b)\tilde \Phi_1({\bf k}_b)<0$, then regions of size $\pi/k_b$ where 
 the volume fraction of the first component is enhanced ($\zeta_1({\bf r})>\zeta_1^0$) 
and at the same time the volume fraction 
of the second component is depleted ($\zeta_2({\bf r})<\zeta_2^0$)
 are followed by regions where  $\zeta_1({\bf r})<\zeta_1^0$ 
and  $\zeta_2({\bf r})>\zeta_2^0$. Thus, when  $\tilde \Phi_2({\bf k}_b)\tilde \Phi_1({\bf k}_b)>0$ 
global (for $k_b=0$) or local  (for $k_b>0$) gas-liquid separation occurs, whereas the  case  
$\tilde \Phi_2({\bf k}_b)\tilde \Phi_1({\bf k}_b)<0$ corresponds to  global or local demixing.

\subsection{Examples}

For an illustration we consider here three simple examples of different types of interaction potentials, 
leading to different behavior.

 {\bf I}. $V^*_{22}=V_{12}^*=0$.

 In this case it follows from Eq.(\ref{i1}) that at the instability with respect to the $k$-mode the temperature is 
\begin{equation}
\label{Tii}
 k_BT=-\frac{\tilde V^*_{11}(k)f^*_{22}}{D},
\end{equation}
and the instability occurs only if $\tilde V^*_{11}(k)<0$. The boundary of stability of the disordered phase 
corresponds to the minimum of $\tilde V^*_{11}(k)$ (maximum of $-\tilde V^*_{11}(k)$). When $V^*_{11}(r)<0$ for all $r$, 
then the minimum of  $\tilde V^*_{11}(k)$
 is assumed for $k=0$, because $\tilde V^*_{11}(0)=\int_{\bf r} V^*_{11}(r)$. However, for potentials that are positive
 for some distances and negative for another distances, like the SALR potential, 
 the minimum of  $\tilde V ^*_{11}(k)$ may be assumed for $k>0$ \cite{ciach:08:1}. In the former case the macroscopic 
separation
 into phases rich- and poor in the first component occurs. In the latter case the instability is induced by the 
concentration wave of the first component with the wavelength $2\pi/k_b$ - 
the regions of excess volume fraction are followed by regions of depleted  volume fraction. 
The volume fraction of the second component is given in Eq.(\ref{r1}). 
For the chosen interactions we have $\tilde C^0_{12}(k)>0$, and the critical mode is  $\tilde \Psi_1({\bf k}_b)$ or
  $\tilde \Psi_2({\bf k}_b)$ for $\tilde C^0_{11}(k)+\tilde C^0_{22}(k)>0$ or 
$\tilde C^0_{11}(k)+\tilde C^0_{22}(k)<0$ respectively 
(see Eqs.(\ref{Cphi})-(\ref{Ceta})).  Note that when $\tilde C_{11}(k_b)=0$ and $\tilde\Phi_2(k_b)=0$, 
there is another instability, with respect to fluctuation of the first component only. The  corresponding temperature,
 $k_BT=-\tilde V_{11}(k_b)/f^*_{11}$, is lower than in Eq.(\ref{Tii}).

 {\bf II.} $V_{12}^*=0$ and $V^*_{22}=V_{11}^*$.

This case corresponds to the simplest interactions for which  $ \tilde U(k)>0$. From Eq.(\ref{i1}) we obtain 
the instability with respect to the $k$-mode
\begin{equation}
  kT=\frac{-2\tilde V^*_{11}(k)}{f^*_{11}+f^*_{22}-\sqrt{(f^*_{11}-f^*_{22})^2+4f^{*2}_{12}}}.
\end{equation}
Again, the boundary of stability of the disordered phase 
corresponds to the minimum of $\tilde V^*_{11}(k)$. Only for $\tilde V^*_{11}(k)<0$ the instability with respect
 to the $k$ mode can occur, because the denominator is positive. For attractive interactions the minimum of
 $\tilde V^*_{11}(k)$ is for $k=0$.
In this case $\tilde C^0_{12}(k)>0$ too, and the critical mode is  $\tilde \Psi_1({\bf k}_b)$ or
  $\tilde \Psi_2({\bf k}_b)$.  In the first case 
$\tilde \Phi_2({\bf k}_b)\tilde \Phi_1({\bf k}_b)<0$ and  in the second case 
$\tilde \Phi_2({\bf k}_b)\tilde \Phi_1({\bf k}_b)>0$, hence demixing and  gas-liquid type separation occurs
 in the first and in the  second case respectively.  The sign of $\tilde C^0_{ii}(k)$ depends on the volume 
fractions and $T$, hence gas-liquid separation and demixing occur for different parts of the phase diagram. 

{\bf III}. $V^*_{22}=V_{11}^*=0$.

In this case  $ \tilde U(k)=-\tilde V^{*2}_{12}(k)<0$ and $\tilde K(k) =-2f^*_{12}\tilde V^{*}_{12}(k)$.
There is one positive and one negative solution of  Eq.(\ref{i1}) for given $k$, and the positive solution takes the form
\begin{eqnarray}
\label{array}
 k_BT=
\left\{ 
\begin{array}{lll}
-\frac{\tilde V^{*}_{12}(k)}{\sqrt{f^*_{11}f^*_{22}} +f^*_{12}} &\;\; {\rm for} &\;\; \tilde V^{*}_{12}(k)<0  \\  
\frac{\tilde V^{*}_{12}(k)}{\sqrt{f^*_{11}f^*_{22}} -f^*_{12}} &\;\; {\rm for} &\;\;\tilde V^{*}_{12}(k)>0
\end{array}
\right.
 \label{T1}.
\end{eqnarray}

It is instructive to consider a particular form of $\tilde V^{*}_{12}(k)$. We choose  square-well  potential 
\begin{equation}
\label{v12}
 V^{*}_{12}(r)=-v \theta(r-1)\theta(a-r)
\end{equation}
where $a>1$ is the range of the potential and $r$ is in $\sigma_{12}$ units. In Fourier representation we have
\begin{equation}
\label{v12F}
\tilde V^{*}_{12}(k)=\frac{4\pi v (ak\cos (ak)-k\cos k+\sin k-\sin(ak))}{k^3}
\end{equation}
\begin{figure}
 \includegraphics[scale=0.4]{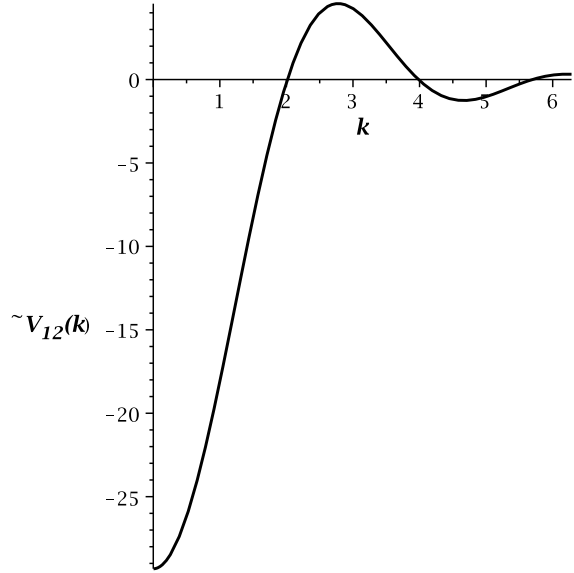}
\caption{ Interaction potential between particles of different kinds in Fourier representation, Eq.(\ref{v12F}) 
for $a=2$. The wave-number $k$ is in $\sigma_{12}^{-1}$ units, the potential is in $v$ units.}
\end{figure}

 $\tilde V^{*}_{12}(k)$ is shown in Fig.3 for $a=2$. For this potential  the MF instability
 is given in Eq.(\ref{T1}), in the upper or in the lower line for $k=0$ or for $k_b\approx 2.78$ 
(the first maximum of $\tilde V^{*}_{12}(k)$) respectively. In the first
 case $\tilde C^0_{12}(0)<0$ and $\tilde \Phi_2(0)\tilde \Phi_1(0)>0$, hence the gas-liquid separation occurs. 
In the  second  case $\tilde C^0_{12}(k_b)>0$ and
 $\tilde \Phi_2(k_b)\tilde \Phi_1(k_b)<0$, hence local demixing occurs. In each case 
$\tilde C^0_{12}(k_b)\tilde \Phi_2(k_b)\tilde \Phi_1(-k_b)<0$
and the Gaussian part of the excess grand potential, Eq.(\ref{OmcoBrG}),
can vanish (in this example $\tilde C^0_{ii}(k)>0$).
The  two surfaces in Eq.(\ref{T1})
 are shown in Figs.4 and 5 for equal sizes of particles, and in Fig.6 for $\delta=0.95$ ($\sigma_1/\sigma_2=39$). 
The period of the most probable inhomogeneities is $2\pi/k_b\approx 2.26\sigma_{12}$. For equal 
sizes $\sigma_{12}=\sigma_1$, but in the case of the large size asymmetry the characteristic extent of inhomogeneities 
is $1.16\sigma_1$. This example is at the boundary of applicability of the mesoscopic description.

For equal sizes gas-liquid separation occurs for low volume fractions, and for higher volume fractions the liquid phase
 undergoes  periodic ordering. The  MF instability with respect to periodic ordering signals tendency for formation 
of an ``ionic`` crystal, where nearest-neighbors are of different kind \cite{ciach:06:0}. 
For large size ratio the phase-separation is found for temperatures lower 
than the temperature at the $\lambda$-surface for all volume fractions. Below the $\lambda$-surface
the system is inhomogeneous.
 Studies beyond MF are necessary to clarify whether  
transition between  gas and lyotropic liquid crystal preempts the gas-liquid separation, or the transition 
between inhomogeneous fluids (containing clusters or other aggregates) occurs. 

\begin{figure}
 \includegraphics[scale=0.35]{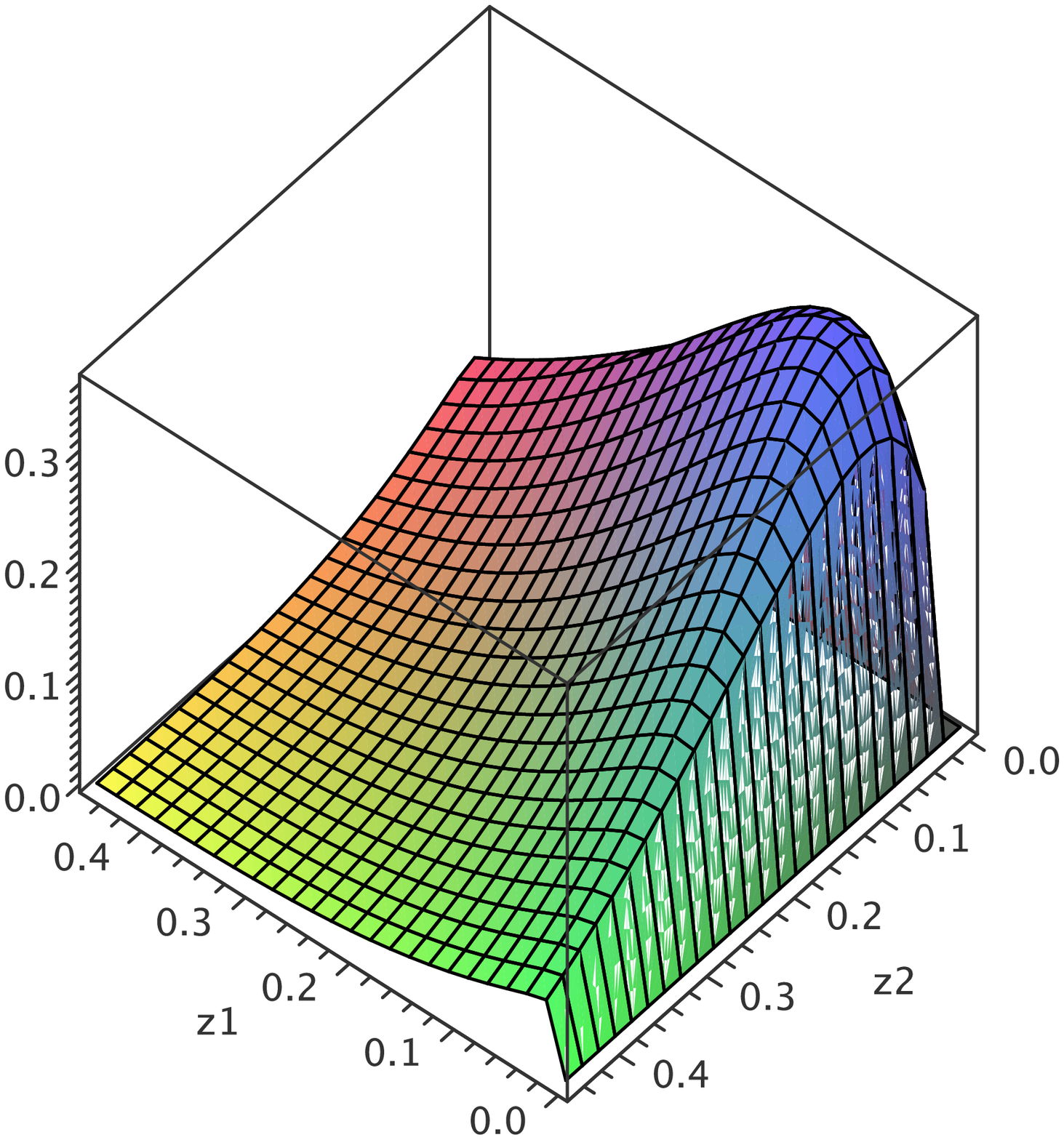}
 \includegraphics[scale=0.35]{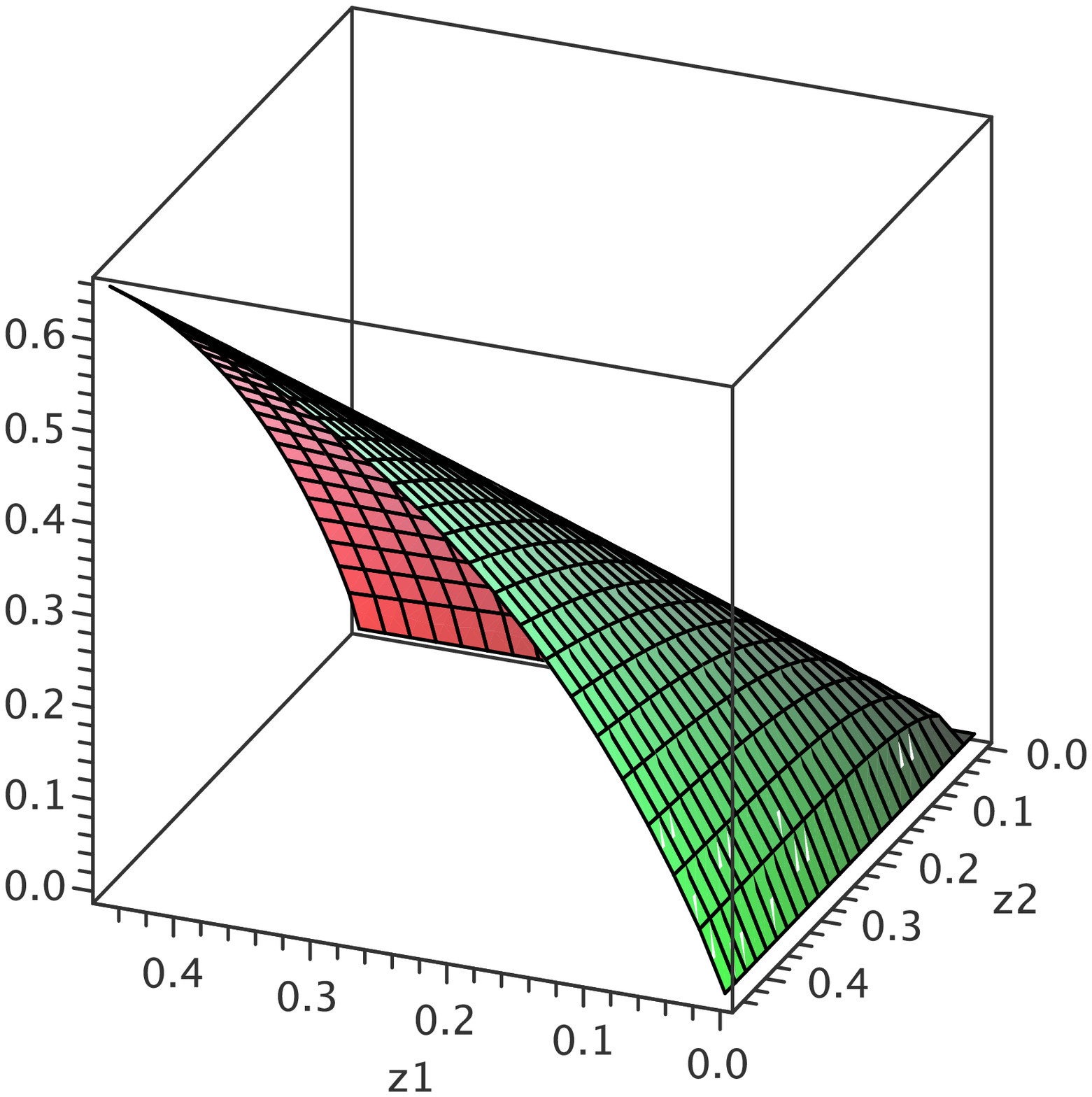}
 \includegraphics[scale=0.4]{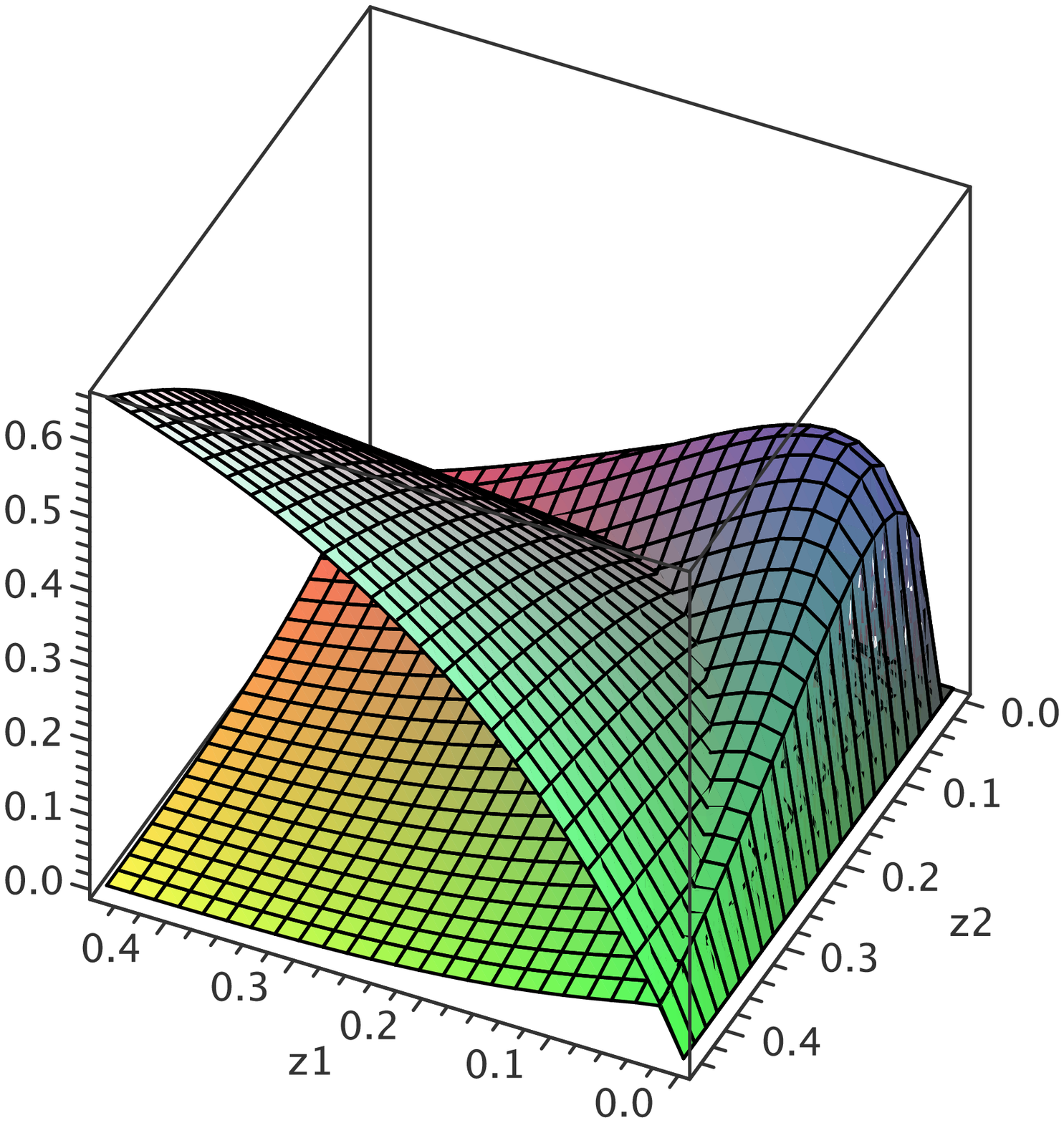}
\caption{Surfaces representing dimensionless temperature $k_BT/v$ at the MF boundary of stability of the homogeneous
 phase for $\tilde V^*_{12}(k)$ given in Eq.(\ref{v12}) with $a=2$, and $\tilde V_{11}(k)= \tilde V_{22}(k)=0$, 
for equal sizes of particles. Instability with respect to gas-liquid separation is shown in top left panel 
and instability with respect to periodic ordering ($\lambda$-surface) in top right panel. 
In bottom panel both surfaces are shown. 
 Volume fractions of the two components 
(dimensionless) here are denoted by $z1$ and $z2$ and in the text by $\zeta_1$, $\zeta_2$ respectively.
 }
\end{figure}
\begin{figure}
 \includegraphics[scale=0.34]{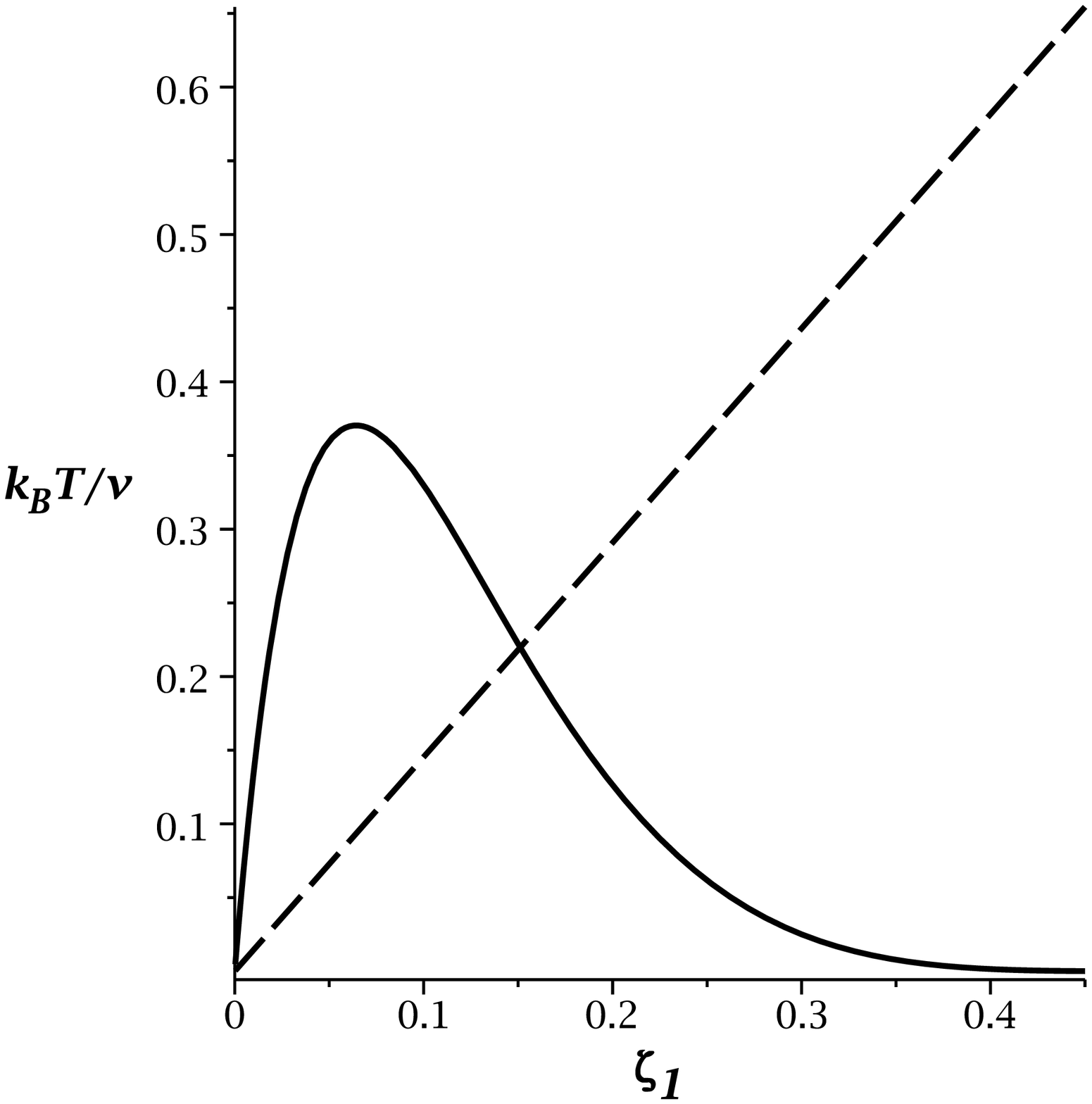} 
\includegraphics[scale=0.34]{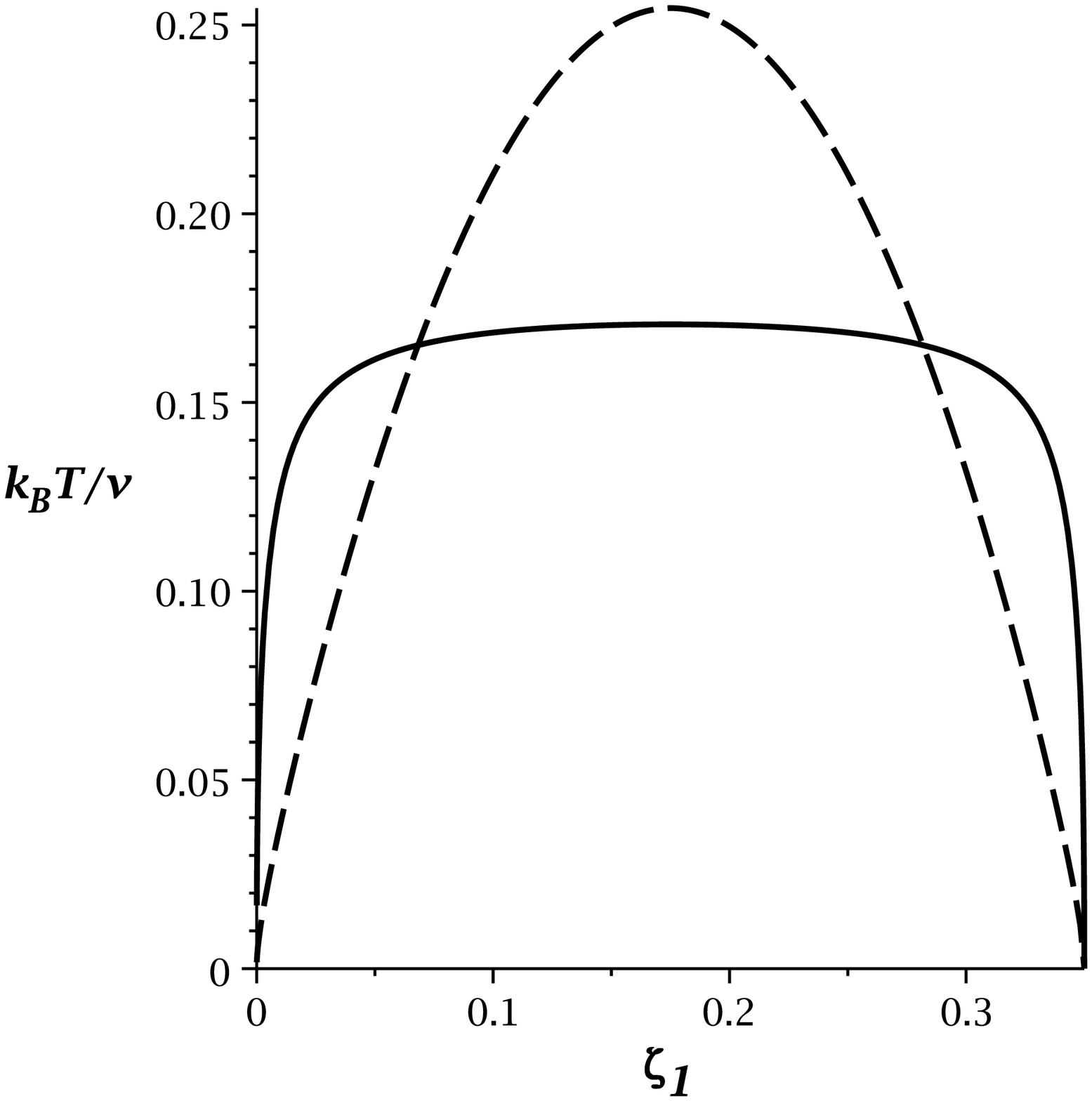}
\caption{Cross-sections through the surfaces shown in Fig.3 (bottom panel). Left: for equal volume fractions of the two 
components, $\zeta_1=\zeta_2$. Right: for  $\zeta_2=0.35-\zeta_1$. Solid and dashed lines
 represent gas-liquid separation and the $\lambda$-surface (the upper and the lower line in Eq.(\ref{T1})) respectively.
 $k_BT/v$ 
and the volume fraction are both dimensionless.}
\end{figure}
\begin{figure}
  \includegraphics[scale=0.4]{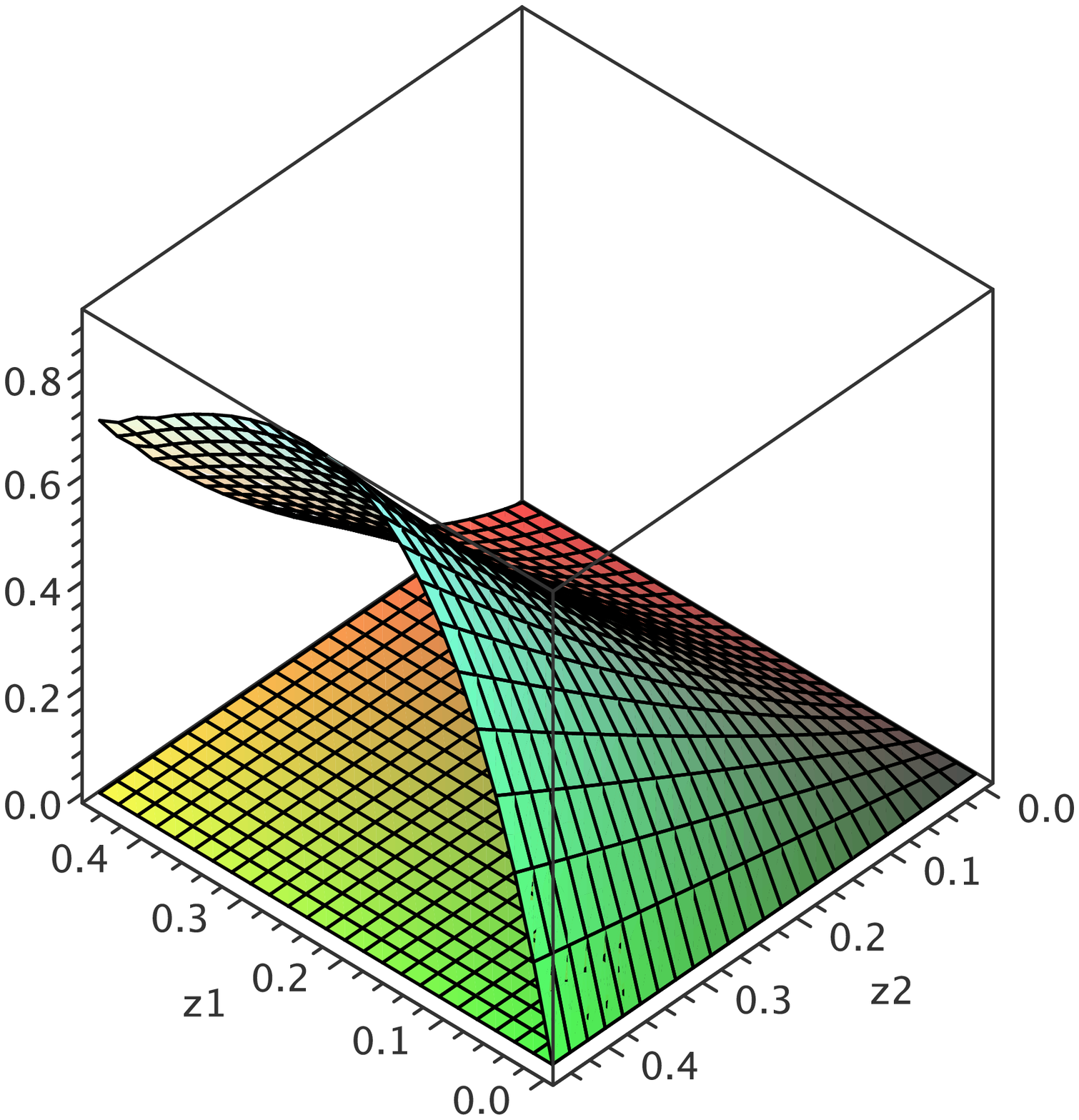}
\caption{ Surfaces representing dimensionless temperature $k_BT/v$ at the MF boundary of stability of the homogeneous 
phase
for $\tilde V^*_{12}(k)$ given in Eq.(\ref{v12}) with $a=2$, and $\tilde V_{11}(k)= \tilde V_{22}(k)=0$, 
 for  $\delta=0.95$ ($\sigma_1/\sigma_2=39$). The gas-liquid separation (lower surface) is metastable in MF 
for all values of the volume fractions. Volume fractions of the two components 
(dimensionless) here are denoted by $z1$ and $z2$ and in the text by $\zeta_1$, $\zeta_2$ respectively.}
\end{figure}

Let us summarize the above examples. When the interaction potentials between like
 particles are attractive for some distances and repulsive for different distances, and  the Fourier 
transform assumes a \textit{negative minimum} for $k>0$, then the system is inhomogeneous below the $\lambda$-surface.
 Such behavior was found already in one-component systems. 
In addition, inhomogeneous
 structures can occur when particles of different kind attract each other and the Fourier transform of the interaction 
potential assumes \textit{positive maximum} for $k>0$. Periodic ordering is enhanced when the size asymmetry increases, 
and the system becomes inhomogeneous even
 for low volume fractions for $\delta>0.9$. Increasing tendency for clustering with increasing size asymmetry was 
observed in ionic systems, in mesoscopic theory \cite{ciach:07:0} and in  simulation studies 
\cite{cheong:03:0,spohr:02:0}.
 
\subsection{Beyond MF stability analysis}

In order to calculate the phase diagram and structure we need to calculate the correlation functions 
beyond MF.
 The explicit expressions for the inverse correlation functions in MF in principle allow for obtaining the 
correlation functions in the Brazovskii-type approximation by self-consistent solutions of Eq.(\ref{CF}).
 In practice this is less trivial than in the one-component case \cite{patsahan:07:0}.

 The grand potential functional of the mesoscopic volume fractions can be obtained once the  matrix 
$\tilde {\bf C}^{co}(k)$
and its eigenvalues are determined from Eqs.(\ref{OmHarval1})-(\ref{OmcoBr}). However, even at the MF 
level determination of the phase diagram is not quite trivial.
The problem simplifies when we restrict our considerations
to phases of particular symmetry. In one-component systems the MF phase diagram was determined for weak ordering in
 Ref.\cite{ciach:08:1,ciach:10:1}, and in two-component case one can proceed in a similar way. The fluctuation
contribution to the grand potential can be obtained from Eqs.(\ref{OmHarval1})-(\ref{OmcoBrG}) and (\ref{CF}).
 We should stress that 
from sec.3 it follows that the phase transitions between inhomogeneous phases are first-order.
 Temperature at the transition is
 shifted compared to the MF result, because the fluctuation contribution to the grand potential in different
 phases is different. 

\section{Summary and discussion}

We developed a mesoscopic DFT for inhomogeneous mixtures in terms of mesoscopic volume fractions.
Local volume
 fraction is suitable for description of inhomogeneous distribution of particles on mesoscopic length scales when 
particles of different species are of significantly different  sizes. This is because
derivation of a Landau-type theory from statistical mechanics requires expansions in local deviations of the volume 
fraction 
(or density) from the average value. Since  volume fractions are less than unity regardless of the size of the
 particles, such expansions are justified from mathematical point of view.  We performed precise coarse-graining by 
introducing the microscopic volume fraction, and then by
 averaging it over mesoscopic regions (see Eq.(\ref{me})) to obtain mesoscopic field. This field is next considered as 
a constraint on the microscopic states. Exact expression for the grand potential consisting of two parts is
 obtained. The first part contains contributions from the microscopic states that are compatible with the
 constraint imposed on the volume fractions on the mesoscopic scale - the mesoscopic volume fraction of each component 
must be equal to the ensemble average. This term has a form
 similar to the grand potential in standard DFT. The second term is the contribution resulting from mesoscopic fluctuations - 
that is, from microscopic states that are not compatible with the ensemble average of the volume fraction on the mesoscopic 
scale. This term in turn has a form similar to the Landau theory. 

In practice additional assumptions and approximations
 are necessary in order to obtain  predictions for particular systems in the framework of this theory. Being interested in inhomogeneities
 on  mesoscopic length scale, we assume that ordering occurs on 
length scales larger than the size of particles, and adopt local density approximation for the hard-sphere reference system.
 The functional can be further simplified under the assumption of weak ordering, by which we mean that local deviations of
 the volume fraction from the space averaged value are small. Under the two assumptions, often valid in soft matter, we
 obtain a functional (Eqs.(\ref{OmHarval1})-(\ref{OmcoBrG}), (\ref{CF}) and (\ref{A})) which is similar to an extension of
 the Landau theory for mixtures combined with simple DFT. The role of the OP's is played by  linear combinations of 
the deviations of the local volume fractions from the
 space-averaged values. Minima of the functional correspond to equilibrium structures. All parameters in the functional are
 expressed in terms of the free-energy density of hard-spheres and in terms of interaction potentials which can have
 arbitrary form. In the case of  one-component systems we obtain LGW or LB theory, when the interaction potential times the
 pair distribution function, $V^{co}$, in Fourier representation is approximated by
\begin{eqnarray}
\label{LL}
\tilde V^{co}(k)=\tilde V^{co}(k_b) +\frac{1}{2}\tilde V^{co}(k_b)^{''}(k-k_b)^{2}.
\end{eqnarray}
 The LGW and LB theories correspond to $k_b=0$ and $k_b>0$ respectively. In original LB theory
 the second term in Eq.(\ref{LL}) is proportional to $(k^2-k_b^2)^2$, but for $k\approx k_b$, i.e. for dominant wave-numbers, 
we have $ (k^2-k_b^2)^2\approx (2k_b)^2(k-k_b)^2$.
 From the theory developed in Ref.\cite{ciach:08:1} either the LGW or the LB theory is obtained as a further 
approximation, depending on the form of the interaction potential. We are not aware of extensions of the phenomenological LB
 theory to mixtures. From the theory developed in this work we can determine  whether the system can separate in homogeneous
 phases, or whether inhomogeneous structures appear on low-temperature side of the $\lambda$-surface. This information can be obtained from the form of interaction
 potentials and from the size ratio of the particles, by performing stability analysis of the homogeneous phase (secs.3 and 4). 
In Landau-type theories separation into homogeneous phases or periodic ordering is an apriori assumption.

Despite strong assumptions and approximations, determination of the phase diagram in this theory is not easy,
 but we can draw some qualitative conclusions already from the relatively simple stability analysis. From the 
approximate form of the inverse correlation function  (Eqs.(\ref{CF}) and (\ref{A})) it follows that instability 
with respect to periodic ordering ($k_b>0$) obtained in MF is removed by mesoscopic fluctuations, as was the case 
also in one component systems. The $\lambda$-surface corresponding to the MF instability may be a borderline between
 homogeneous and inhomogeneous structure of the disordered phase. For high volume fractions formation of some kind 
of periodic crystal can be expected on this side of the $\lambda$-surface which corresponds to inhomogeneous structure. 
For low volume fractions we may expect formation of ordered clusters, where domains with increased and depleted volume 
fraction of particular components are periodically ordered. However, the clusters ('living polymers') can have different sizes and locations in space 
as a result of the mesoscopic fluctuations. Simple examples that illustrate the theory for two components indicate that
 formation of inhomogeneous distribution of particles is enhanced when the size ratio between the particles of different
 components increases. Further work is necessary for determination of phase diagrams for particular systems within the 
framework of the theory developed in this work, and for verification of validity of our approximations in different 
experimental systems. It is important to note that  formation (or not) of inhomogeneous distribution of particles in mixtures
 depends not only on the interaction potentials, but on a combined effect of interactions and size ratios. 
In future studies predictions of this theory for two components should be compared with predictions of the theory 
in which only the big particles are taken into account explicitly, and the small components lead only to solvent-mediated
 effective interactions. 

{\bf Acknowledgments}
This work is dedicated to Prof. Robert Evans on the occasion of his birthday.
I would like to thank Dr. Oksana Patsahan for fruitful discussions. 
Partial supports by the Polish Ministry of Science and Higher Education, Grant No NN 202 006034,
 and  by the Ukrainian-Polish joint research project 
``Statistical theory of complex systems with electrostatic interactions'' are gratefully acknowledged. 

\section{Appendix}

The partial derivatives of the dimensionless  free-energy density for  hard spheres,
 $f^*_{\alpha\beta}(\zeta_1,\zeta_2)=\sigma_{12}^3\beta f^h_{\alpha\beta}(\zeta_1,\zeta_2)$ are obtained from
the explicit expressions for the chemical potentials in the
hard-sphere mixture derived in Ref.\cite{lebowitz:64:0}. Direct differentiations lead to the 
following expressions
\begin{eqnarray}
\label{a++}
f^*_{\alpha\alpha}(\zeta_1,\zeta_2)=
\frac{6}{\pi r_{\alpha}^3}\Bigg[\frac{1}{\zeta_{\alpha}} +
\frac{8}{1-\zeta}+\frac{15r_{\alpha}X_1+6r_{\alpha}^2X_2+r_{\alpha}^3X_3}{(1-\zeta)^2}
\\
\nonumber
+\frac{18r_{\alpha}^2X_1^2+6r_{\alpha}^3X_1X_2}{(1-\zeta)^3}+\frac{9r_{\alpha}^3X_1^3}{(1-\zeta)^4}\Bigg]
\end{eqnarray}
and
\begin{eqnarray}
\label{a+-}
f^*_{12}(\zeta_1,\zeta_2)=\frac{6}{\pi( r_1r_2)^3}\Bigg[\frac{8}{1-\zeta}+
\frac{2r_1r_2(6+r_1r_2)X_1+8r_1^2r_2^2X_2}{(1-\zeta)^2}
\\
\nonumber
+
\frac{18r_1^2r_2^2X_1^2+6r_1^3r_2^3X_1X_2}{(1-\zeta)^3}+
\frac{9r_1^3r_2^3X_1^3}{(1-\zeta)^4}\Bigg]
\end{eqnarray}
where
\begin{equation}
X_n=\frac{\zeta_1}{r_1^n}+\frac{\zeta_2}{r_2^n}
\end{equation}
and $r_{\alpha}$ is defined in Eq.(\ref{rpm})


\begin{thebibliography}{39}
\bibitem{hansen:76:0}
J. Hansen and I. McDonald, \emph{Theory of Simple Liquids}   (Academic Press,
  London, 1976).

\bibitem{zinn-justin:89:0}
Zinn-Justin, \emph{Quantum Field Theory and Critical Phenomena}   (Clarendon
  Press, Oxford, 1989).

\bibitem{amit:84:0}
D.J. Amit, \emph{Field Theory, the Renormalization Group and Critical
  Phenomena}   (World Scientific, Singapore, 1984).

\bibitem{yukhnovskii:58:0}
I. Yukhnovskii, Sov.Phys. JETP \textbf{34}, 263 (1958).

\bibitem{yukhnovskii:78:0}
I.R. Yukhnovskii, \emph{Phase Transitions of the Second Order, Collective
  Variable Methods}   (World Scientific, Singapore, 1978).

\bibitem{caillol:06:0}
J.M. Caillol, O. Patsahan and I. Mryglod, {\it Physica A} \textbf{368}, 326
  (2006).

\bibitem{parola:85:0}
A. Parola and L. Reatto, {\it Phys. Rev. A} \textbf{31}, 3309 (1985).

\bibitem{parola:95:0}
A. Parola and L. Reatto, {\it Adv. Phys.} \textbf{44}, 211 (1995).

\bibitem{patsahan:99:0}
O. Patsahan, Physica A \textbf{272}, 358 (1999).

\bibitem{pini:00:0}
D. Pini, J. Ge, A. Parola and L. Reatto, Chem. Phys. Lett. \textbf{327}, 209
  (2000).

\bibitem{stradner:04:0}
A. Stradner, H. Sedgwick, F. Cardinaux, W. Poon, S. Egelhaaf and P.
  Schurtenberger, {\it Nature} \textbf{432}, 492 (2004).

\bibitem{campbell:05:0}
A.I. Campbell, V. J.Anderson, J.S. van Duijneveldt and P. Bartlett, {\it Phys.
  Rev. Lett.} \textbf{94}, 208301 (2005).

\bibitem{archer:07:0}
A.J. Archer, D. Pini, R. Evans and L. Reatto, {\it J. Chem. Phys.}
  \textbf{126}, 014104 (2007).

\bibitem{archer:07:1}
A.J. Archer and N.B. Wilding, {\it Phys. Rev. E} \textbf{76}, 031501 (2007),
  and references therein.

\bibitem{pini:06:0}
D. Pini, A. Parola and L. Reatto, {\it J. Phys.:Cond. Mat.} \textbf{18}, S2305
  (2006).

\bibitem{imperio:04:0}
A. Imperio and L. Reatto, {\it J. Phys.:Cond. Mat.} \textbf{16}, 3769 (2004).

\bibitem{brazovskii:75:0}
S.A. Brazovskii, {\it Sov. Phys. JETP} \textbf{41}, 8 (1975).

\bibitem{leibler:80:0}
L. Leibler, {\it Macromolecules} \textbf{13}, 1602 (1980).

\bibitem{teubner:87:0}
M. Teubner and R. Strey, {\it J. Chem. Phys.} \textbf{87} (5), 3195--3200
  (1987).

\bibitem{fredrickson:87:0}
G.H. Fredrickson and E. Helfand, {\it J. Chem. Phys.} \textbf{87}, 67 (1987).

\bibitem{gompper:94:0}
G. Gompper and M. Schick, {\it Phys. Rev. E} \textbf{49} (2), 1478--1482
  (1994).

\bibitem{podneks:96:0}
V.E. Podneks and I.W. Hamley, {\it Pis'ma Zh. Exp. Teor. Fiz.} \textbf{64}, 564
  (1996).

\bibitem{ciach:01:2}
A. Ciach and W.T. G\'o\'zd\'z, {\it Annu. Rep.Prog. Chem., Sect.C} \textbf{97},
  269 (2001), and references therein.

\bibitem{evans:79:0}
R. Evans, {\it Adv. Phys.} \textbf{28}, 143 (1979).

\bibitem{ciach:08:1}
A. Ciach, {\it Phys. Rev. E} \textbf{78}, 061505 (2008).

\bibitem{dijkstra:99:0}
M. Dijkstra, J.M. Brader and R. Evans, {\it J. Phys.:Cond. Mat.} \textbf{11},
  10079 (1999).

\bibitem{dijkstra:00:0}
M. Dijkstra, R. van Roij and R. Evans, {\it J. Chem. Phys.} \textbf{113}, 4799
  (2000).

\bibitem{ciach:06:2}
A. Ciach and O. Patsahan, {\it Phys. Rev. E} \textbf{74}, 021508 (2006).

\bibitem{ciach:00:0}
A. Ciach and G. Stell, {\it J. Mol. Liq.} \textbf{87}, 255 (2000).

\bibitem{ciach:03:1}
A. Ciach, W.T. G\'o\'zd\'z and R.Evans, {\it J. Chem. Phys.} \textbf{118}, 3702
  (2003).

\bibitem{ciach:06:1}
A. Ciach, {\it Phys. Rev. E} \textbf{73}, 066110 (2006).

\bibitem{patsahan:07:0}
O. Patsahan and A. Ciach, {\it J. Phys.: Condens. Matter} \textbf{19}, 236203
  (2007).

\bibitem{lebowitz:64:0}
J.L. Lebowitz and J.S. Rowlinson, {\it J. Chem. Phys.} \textbf{41}, 133 (1964).

\bibitem{ciach:07:0}
A. Ciach, W.T. G\'o\'zd\'z and G. Stell, {\it Phys. Rev. E} \textbf{75}, 051505
  (2007).

\bibitem{patsahan:10:0}
O. Patsahan and T. Patsahan, {\it Phys. Rev. E} \textbf{81}, 031110 (2010).

\bibitem{ciach:06:0}
A. Ciach, W.T. G\'o\'zd\'z and G. Stell, {\it J. Phys. Cond. Mat.} \textbf{18},
  1629 (2006).

\bibitem{cheong:03:0}
D. Cheong and A. Panagiotopoulos, {\it J. Chem. Phys.} \textbf{119}, 8526
  (2003).

\bibitem{spohr:02:0}
E. Spohr, B. Hribar and V. Vlachy, {\it J. Phys. Chem B.} \textbf{106}, 2343
  (2002).

\bibitem{ciach:10:1}
A. Ciach and W.T. G\'o\'zd\'z, {\it Condensed Matter Physics} \textbf{13},
  23603 (2010).

\end{thebibliography}
\end{document}